\documentclass[aps,prb,twocolumn,superscriptaddress,amsmath,amssymb]{revtex4-1}
\newcommand{\bra}[1]{\langle #1|}
\newcommand{\ket}[1]{|#1\rangle}
\newcommand{\braket}[2]{\langle #1|#2\rangle}
\newcommand{\kk}{\mathbf{k}}
\newcommand{\qq}{\mathbf{q}}
\newcommand{\QQ}{\mathbf{Q}}
\newcommand{\RR}{\mathbf{R}}
\usepackage{xcolor}
\usepackage{graphicx}
\usepackage{bm}
\usepackage{lineno}
\usepackage{comment}
\usepackage{soul}

\begin{document}

\title{Direct and indirect excitons in boron nitride polymorphs:\\
a story of atomic configuration and electronic correlation}

\author{Lorenzo Sponza}
\affiliation{Laboratoire d'Etude des Microstructures, ONERA-CNRS, UMR104, Universit\'e Paris-Saclay, BP 72, 92322 Ch\^atillon Cedex, France}
\author{Hakim Amara}
\affiliation{Laboratoire d'Etude des Microstructures, ONERA-CNRS, UMR104, Universit\'e Paris-Saclay, BP 72, 92322 Ch\^atillon Cedex, France}
\author{Claudio Attaccalite}
\affiliation{Aix Marseille University and CNRS, CINAM, UMR7325, Campus de Luminy, case 913, 13288 Marseille, France}
\affiliation{University of Rome Tor Vergata, Rome, Italy}
\author{Sylvain Latil}
\affiliation{CEA, IRAMIS, SPEC, GMT, 91191 Gif-sur-Yvette, France}
\author{Thomas Galvani}
\affiliation{Physics and Materials Science Research Unit, University of Luxembourg, 162a avenue de la Fa\"iencerie, L-1511 Luxembourg, Luxembourg, EU}
\author{Fulvio Paleari}
\affiliation{Physics and Materials Science Research Unit, University of Luxembourg, 162a avenue de la Fa\"iencerie, L-1511 Luxembourg, Luxembourg, EU}
\author{Ludger Wirtz}
\affiliation{Physics and Materials Science Research Unit, University of Luxembourg, 162a avenue de la Fa\"iencerie, L-1511 Luxembourg, Luxembourg, EU}
\author{Fran\c{c}ois Ducastelle}
\affiliation{Laboratoire d'Etude des Microstructures, ONERA-CNRS, UMR104, Universit\'e Paris-Saclay, BP 72, 92322 Ch\^atillon Cedex, France}

\begin{abstract}
We compute and discuss the electronic band structure and excitonic dispersion of hexagonal boron nitride (hBN) in the single layer configuration and in three bulk polymorphs (usual AA' stacking, Bernal AB, and rhombohedral ABC). We focus on the changes in the electronic band structure and the exciton dispersion induced by the atomic configuration and the electron-hole interaction. Calculations are carried out on the level of \textit{ab initio} many-body perturbation theory (GW and Bethe Salpeter equation) and by means of an appropriate tight-binding model.
We confirm the change from direct to indirect electronic gap when going from single layer to bulk systems and we give a detailed account of its origin by comparing the effect of different stacking sequences.
We emphasize that the inclusion of the electron-hole interaction is crucial for the correct description of the momentum-dependent dispersion of the excitations. It flattens the exciton dispersion with respect to the one obtained from the dispersion of excitations in the independent-particle picture. In the AB stacking this effect is particularly important as the lowest-lying exciton is predicted to be direct despite the indirect electronic band gap.
\end{abstract}

\date{\today}

\maketitle

%%%%%%%%%%%%%%%%%%%%%%%%%%%%%%%%%%%%%%%%%%%%%%%%%%%%%%%%%%%

\section{Introduction}

Hexagonal boron nitride (hBN) is a wide band gap semiconductor which makes it the object of growing scientific interest. In the field of optics, hBN is known to exhibit a natural hyperbolic behaviour\cite{Jacob2014} and single photon emission properties.\cite{Tran2016,Bourrellier2016,Wu2016} It is also an efficient emittor in the UV range because of its optical gap of about 6 eV.\cite{Watanabe2009b} Many recent experimental\cite{Watanabe2006,Jaffrennou2007,Watanabe2009a,Museur2011,Pierret2014,bourrellier_acsphot2014,Du2015,Cassabois2016a,Cassabois2016b,Doan2016,Li2016,Henck2017,Schue2016,Schue2017,Vuong2017} and theoretical studies\cite{Blase1995,Arnaud2006,Arnaud2008,Wirtz2006,Wirtz2008,bourrellier_acsphot2014,aggoune_prb2018} concerning its electronic structure and optical properties have provided new elements to understand the peculiar properties of hBN, namely the explanation of the high intensity of the luminescence signal in the bulk despite its indirect band gap, or of the so-called Stokes shift between absorption and luminescence.\cite{Schue2018}

All first principles calculations predict that the electronic gap is direct for the single layer,\cite{Ribeiro2011,Galvani2016} but indirect in the bulk, and actually even for a bilayer.\cite{Kang2016,Paleari2018} These results have been obtained within effective independent particle schemes like the density functional theory or the GW quasiparticle equation which includes only electron-electron correlations. However, when discussing two-particle response functions (optical spectroscopy, energy loss spectroscopy or X-ray scattering), these schemes are often insufficient because they lack important electron-hole interactions. This is particularly true in systems where the screening is weak, like in thin films and large gap bulk semiconductors, which is the case for hBN. A key element to gain this insight in hBN has been the analysis of the excitonic dispersion at finite $\bm{Q}$ along the $\Gamma$M and the $\Gamma$K lines.\cite{Sponza2018}

\begin{figure}
\centering
\includegraphics[width=0.48\textwidth]{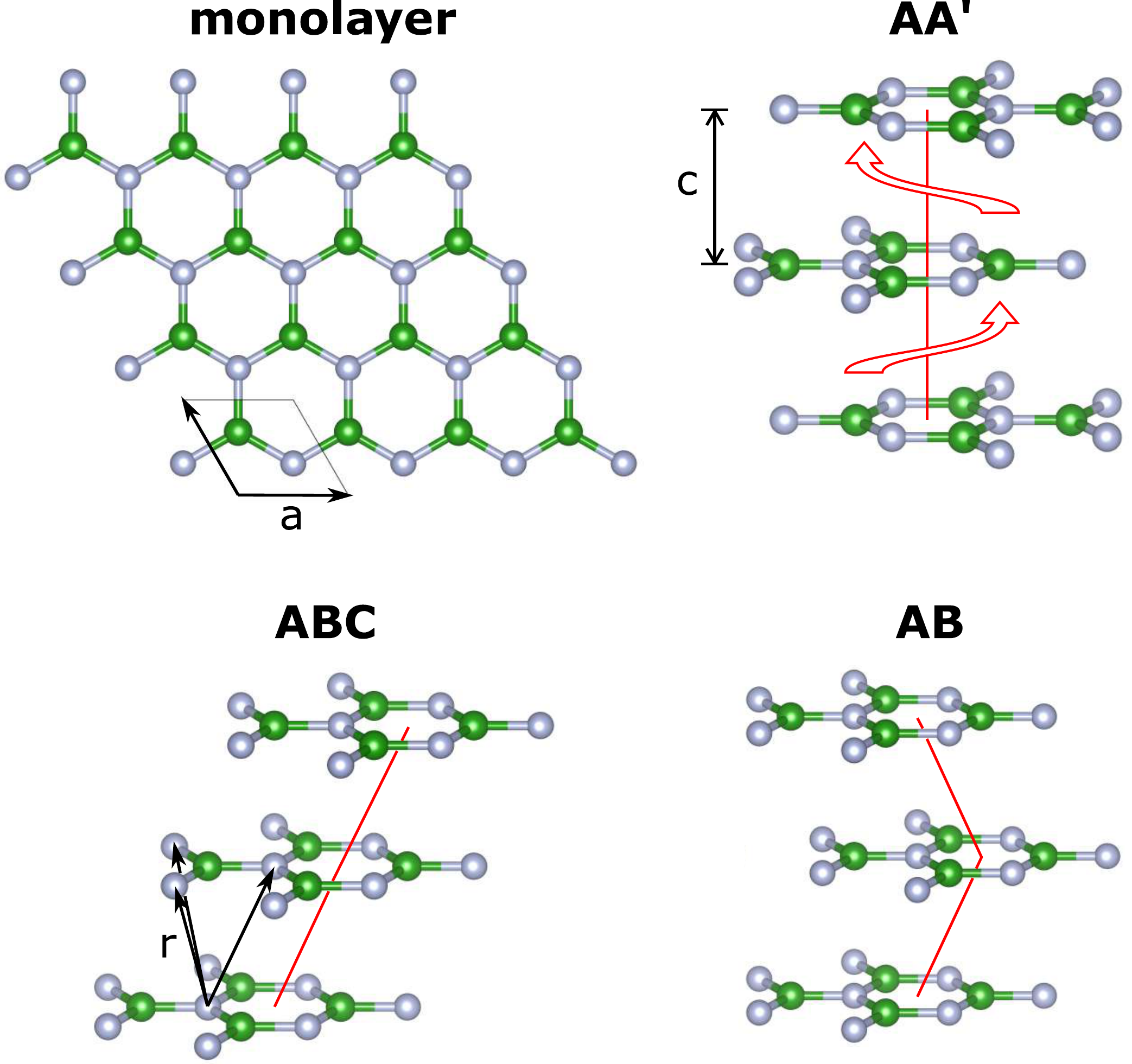}
\caption{The atomic structure of hBN monolayer, the AA', the AB and the ABC stacking. In the latter, the unitary lattice vectors of the rhombohedral cell are also reported. A red line joining the centers of three B$_3$N$_3$ hexagons highlights the difference between the three bulk structures.}
\label{fig:structure}
\end{figure}

In this article we push the analysis further, discussing in detail the dispersion of the excitons especially in the $\Gamma$K direction, relevant for optical spectroscopy, with the intent of tracking down the structural elements affecting the excitonic properties and the excitonic dispersi1on in particular. To this aim, the case of hBN monolayer is investigated and compared with three different polymorphs of bulk hBN, reported in Figure~\ref{fig:structure}. The usual structure is the so-called AA' stacking where B and N atoms alternate along the stacking axis. Another stable structure, although less common, is the ABC rhombohedral one,\cite{Chubarov2012} and finally the AB Bernal stacking has been reported for few layers.\cite{Kim2013} Our analysis combines \textit{ab initio} Bethe-Salpeter calculations\cite{Martin2016} and a tight-binding Wannier model which has already been shown to be fairly accurate for hBN.\cite{Galvani2016}
Our work integrates and completes the discussion of excitonic effects in multilayer hBN at $\bm{Q} = 0$ published elsewhere,\cite{Paleari2018} as well as the investigation of optical spectra in different stacking sequences.\cite{bourrellier_acsphot2014,aggoune_prb2018}

The paper is organized as follows. In section II we detail the computational parameters used for the \textit{ab initio} calculations and we devise the tight-binding (TB) model used. In section III we focus on the monolayer calculation. By comparing the \textit{ab initio} and the TB results, we are able to validate the approach and to appreciate the reliability of the TB predictions in this system. Moreover, the monolayer hosts the fundamental in-plane physics and constitutes the building block of the three bulk polymorphs. In section IV, the comparison of the different bulk phases is carried out with a highlight on the way the stacking sequence affects the reference in-plane electronic and excitonic properties. Finally, in section V we draw our conclusions.

\section{Theoretical methods }

In this section we report the parameters used in the \textit{ab-initio} calculations and introduce the tight-binding model developed for the calculation of one-particle and two-particle properties of the four systems considered.

\subsection{Ab initio simulations}

\subsubsection{Free-standing single layer}

For the boron nitride single layer we use an in-plane lattice parameter of  $a=2.50$~\AA~and a distance between the periodic replica of 13~\AA. The Kohn-Sham states and energies, the GW corrections, and the excitonic properties have been computed with the GPAW code.\cite{Enkovaara2010} Projector-augmented wave (PAW) methods have been used for both atomic species. DFT energies and wave functions have been obtained within the PBE exchange-correlation potential, using a plane-wave cutoff energy of 40~Ha. We did not find relevant differences in the exciton dispersion with respect to the local density approximation (LDA) results. For the density calculation we used a $12\times 12 \times 1$ $\Gamma$-centred grid. The Bethe-Salpeter equation (BSE) has been solved using a truncated Coulomb potential on a $36\times 36 \times 1$ $\Gamma$-centred k-point grid, with a cut-off energy of 100~eV and including 60 bands in the calculation of the dielectric constant. Three valence and three conduction bands have been included in the excitonic Hamiltonian and quasiparticle energies have been approximated with a scissor operator of 2.75 eV adjusted on a recently published GW result.\cite{Galvani2016}

\subsubsection{Bulk structures}
\label{sec:methods:ab-initio:bulk}

For the three bulk structures we used the same in-plane parameter $a$ as for the monolayer and an interlayer distance $c=3.25$~\AA. The value of $c$ is in agreement with measures of the AA' stacking published in a previous work of ours.\cite{Fossard2017}
The Kohn-Sham equations and the GW corrections have been calculated  with the plane-wave simulation package ABINIT.\cite{abinit} Norm-conserving Troullier-Martins pseudopotentials have been used for both atomic species. DFT energies and wave functions have been obtained within the local density approximation (LDA) to the exchange-correlation potential, using a plane-wave cutoff energy of 30~Ha for the three stackings. The hexagonal Brillouin zone of the AA' and the AB stackings have been sampled with a $8\times 8\times 4 \; k$-point grid, while the Brillouin zone of the rhombohedral cell of the ABC stacking has been sampled with a $9\times 9 \times 9$ grid. All $k$-point grids are $\Gamma$-centred.

Quasiparticle corrections have been obtained within the perturbative G$_0$W$_0$ approach. They have been computed on all points of a  $6\times 6\times 4$ $\Gamma$-centred grid for AA' and AB stacking and on a $9 \times 9 \times 9$ $\Gamma$-centered grid for ABC. A cutoff energy of 30~Ha defines the matrix dimension and the basis of wave functions for the calculation of the exchange part of the self-energy. The correlation part has been computed including 600 and 150 bands in hexagonal and rhombohedral structures respectively, and using the same wave function basis as for the exchange part. To model the dielectric function, the contour deformation method has been used for AA', computing the dielectric function up to 60~eV, but this showed to give negligible improvements with respect to the Godby-Needs plasmon pole approximation, so the latter has been used in the other structures. To obtain the GW energies along high-symmetry lines and on finer grids, the GW corrections have been interpolated. Note that the Brillouin zone of the ABC rhombohedral stacking is larger than the hexagonal cell of AA' and AB systems, so the band structure has been folded to the hexagonal cell for a consistent comparison. For a given point on the hexagonal cell, three points in the rhombohedral cell have been taken: $\kk$ and $\kk \pm \tilde{\kk}$ with $\tilde{\kk} = \left( 1/3 , 1/3 , 1/3 \right)$ expressed in the reciprocal coordinates of the rhombohedral cell.

The macroscopic dielectric function $\epsilon_M(\bm{q},\omega)$ has been calculated on the GW-BSE level using the EXC code.\cite{exc} For the hexagonal structures AA' and AB, we included six valence bands and three conduction bands, fixing a cut-off energy of 360~eV for both the matrix dimension and the wave function basis. The static dielectric matrix entering  the BSE kernel has been computed within the random phase approximation with local fields, including 350 bands and with cut-off energies of 120~eV and 200~eV for the matrix dimension and the wave function basis respectively. In the case of the ABC structure, the BSE has been solved including four valence and four conduction bands and keeping the same cut-off as for the hexagonal structures. The static screening used has the same parameters as for the underlying GW calculation. With these parameters, the energies  of the excitons are converged within 0.05~eV in all structures. The dispersion of the exciton as a function of $\qq$ in the ABC cell required the same folding procedure explained above.

\subsection{Tight-binding model}

\subsubsection{Free-standing single layer}

\begin{figure}
\centering
\includegraphics[width=0.45\textwidth]{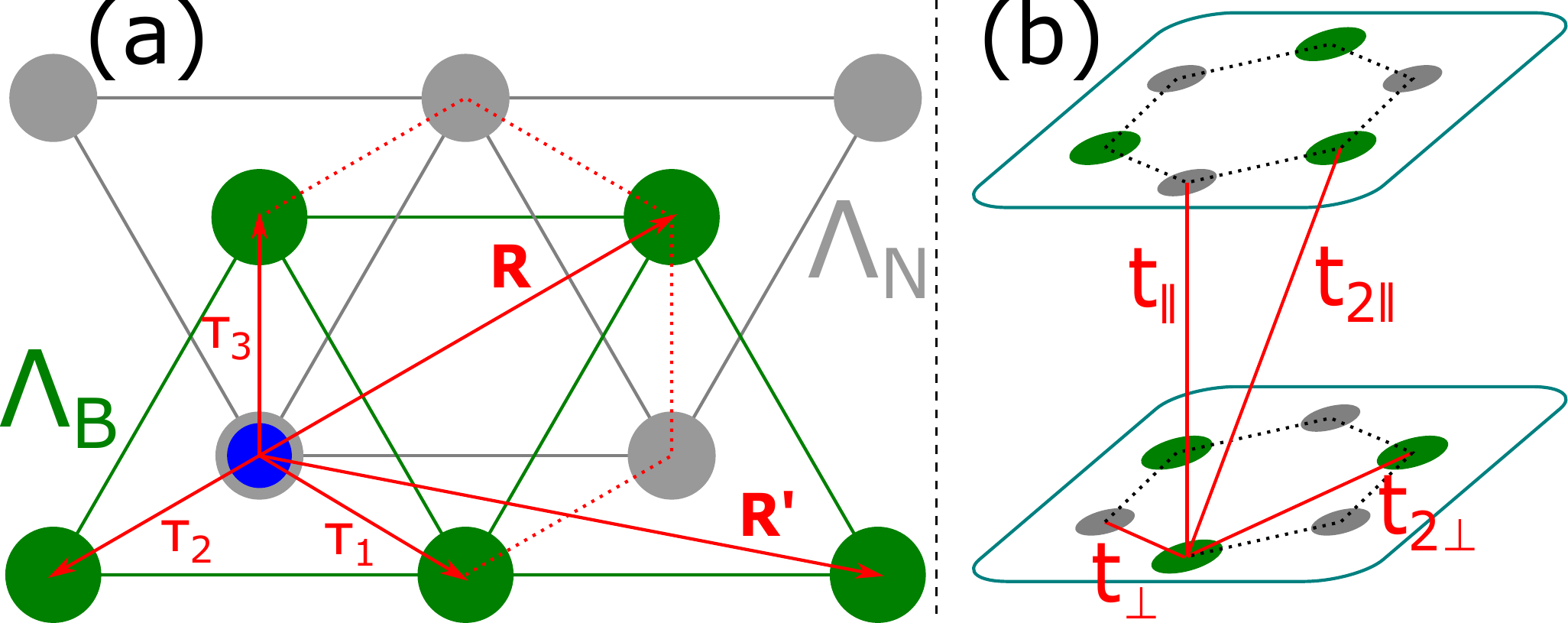}
\caption{(a) The two triangular sublattices $\Lambda_B$ (green) and $\Lambda_N$ (grey) forming the hBN honeycomb lattice. Red arrows mark the $1nn$ vectors $\bm{\tau}_\alpha$ and the positions $\RR$, $\RR'$ of the hopping electron when the hole is placed on an N site at the origin (blue circle).
(b) Scheme of the four hopping terms $t_\perp$, $t_{2\perp}$, $t_\parallel$, and $t_{2\parallel}$ in the AA' stacking.}
\label{fig:triangular_lattices}
\end{figure}

We first consider the TB model of the free-standing monolayer, already introduced by some of us.~\cite{Galvani2016} Let us recall its main features. The real space honeycomb lattice can be divided into two triangular sublattices $\Lambda_B$ and $\Lambda_N$, either connecting all sites of B or N atoms respectively. Vectors $\bm{\tau}_{\{1,2,3\}}$ connect neighbouring sites of the two sublattices. A scheme of the structure is reported in Figure~\ref{fig:triangular_lattices}(a). From the energetic point of view, we note $\pm\Delta$ the on-site energies on B (+) and N ($-$). First- and second-nearest-neighbour (from now on contracted in \textit{1nn} and \textit{2nn}) hopping integrals are $t_\perp$ and $t_{2\perp}$ respectively. The latter is assumed to be equal for $B-B$ and $N-N$ hopping. With these ingredients and by passing to the Bloch representation as in Ref. [\onlinecite{Galvani2016}], the TB energies in the single layer can be approximated  for $\kk_\perp$  along the $KM$ direction as:
\begin{equation}
\begin{split}
\lefteqn{E_{\kk_\perp e(h)} \approx} \\
& \approx  \pm\underbrace{\left(\Delta + \frac{|t_\perp \gamma(\kk_\perp)|^2}{2\Delta} \right)}_\textrm{first neighbours} + \underbrace{t_{2\perp}\left(|\gamma(\kk_\perp)|^2-3\right)}_\textrm{second neighbours}\  \;  ,
\end{split}
\label{eq:TB_single-layer_1particle}
\end{equation}
where $\gamma(\kk_\perp) = \sum_{\alpha=1,2,3} e^{i\kk_\perp \bm{\tau}_\alpha}$ and $\kk_\perp$ is strictly in-plane.  
In the above expression the (+) sign is for conduction $e$ states localized on $\Lambda_B$ sites, and ($-$) is for valence $h$ states localized on $\Lambda_N$. Note that without the \textit{2nn} contribution valence and conduction bands would be symmetric.

%& =   \pm\underbrace{\sqrt{ \Delta^2 + |t_\perp \gamma(\kk_\perp)|^2} }_\textrm{first neighbours} + \underbrace{t_{2\perp}\left(|\gamma(\kk_\perp)|^2-3\right)}_\textrm{second neighbours}\\

\begin{figure*}
\centering
\includegraphics[height=5cm]{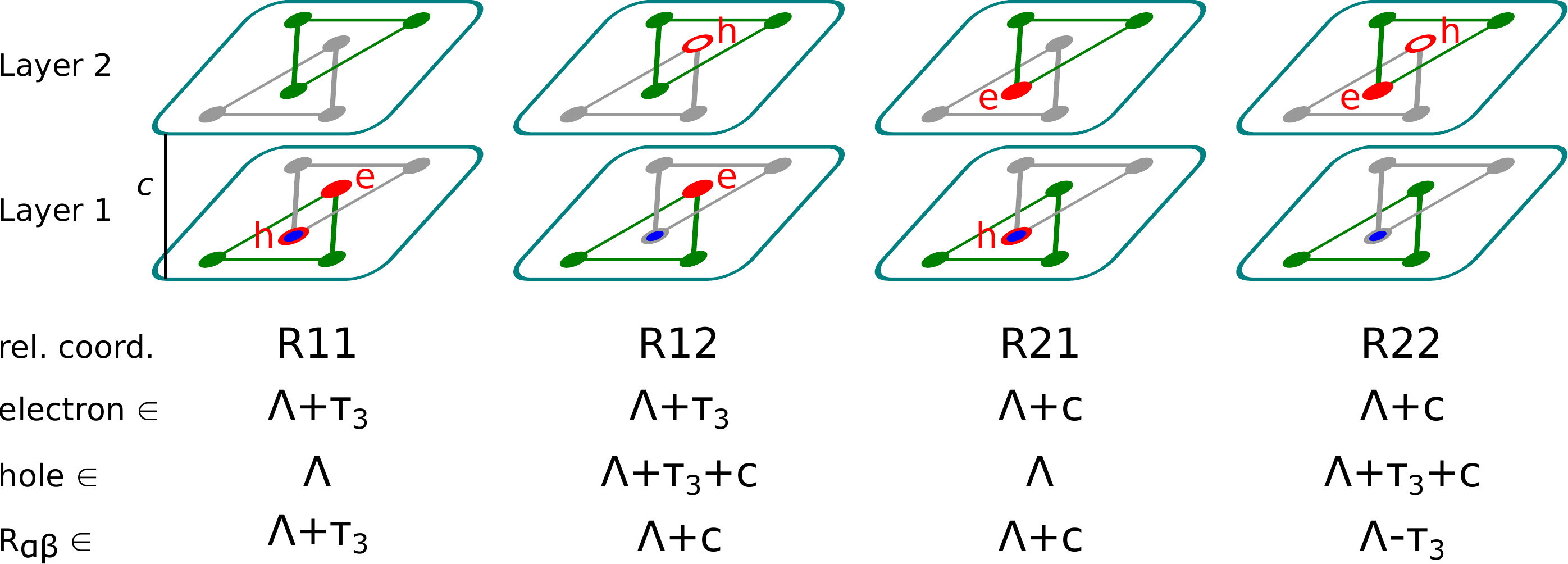}
\caption{The four configurations $\RR_{\alpha,\beta}$ of electron-hole pairs in AA' stacking. The origin is marked with a blue spot on the $\Lambda_N$ sublattice of layer 1. Electrons and holes localize on different $\Lambda + \bm{u}$ sublattices with $\bm{u}$ appropriate translation vector ($\Lambda$ is defined as $\Lambda_N$ on layer 1, and $\Lambda_B=\Lambda+\bm{\tau}_3$ on layer 1)\cite{Note1}. The relative distance $\RR_{\alpha,\beta}$ belongs also to a shifted lattice $\Lambda+\bm{u}$.}
\label{fig:relative-coordinates_tb}
\end{figure*}

The description of the exciton relies on a Wannier TB model,\cite{Galvani2016} but in this work we extend it beyond the optical limit and add improvements on the electron-hole interaction. Let us first recall the basics of the model in the optical limit $\QQ=0$. The fact that we can describe the concerned one-electron $\pi$ bands with atomic-like Wannier functions allows us to work directly with excitonic Wannier equations in real space.\cite{Wannier1937,Knox1963,Toyozawa2003,Bechstedt2015,Martin2016} Assuming the hole to be fixed on a N site, and using relative coordinates $\RR$ for the electron-hole pair, we have reduced the problem to the one of the electron hopping on sites of $\Lambda_B$ in the presence of an attractive impurity located at the origin, as represented schematically in Figure~\ref{fig:triangular_lattices}(a). The corresponding Bethe-Salpeter-like Hamiltonian $H_{eh}=H^{0}_{eh}+\Xi+V$ contains a kinetic part $H^0_{eh}$, a screened Coulomb term $\Xi$ that can be taken as a fitting parameter and an exchange term $V\propto\QQ$ that actually vanishes in the $\QQ=0$ limit.

We extend now the model to $\QQ\neq 0$ hence allowing the exciton to move across the layer. Through the definition of  appropriate Bloch states $\ket{\RR,\QQ}$ defined in Appendix \ref{TBdisp}, the momentum $\QQ$ is a good quantum number related to the propagation of the center of mass of the electron-hole pair. We can then adopt a mixed representation, where the motion of the pair is treated in reciprocal space $\QQ$ and the relative electron-hole distance in real space $\RR$. For the monolayer, it can be demonstrated that the simple model above can be extended to a Hamiltonian $H_{eh} = H^0_{eh} + \Xi + V$
where the kinetic part reads:
\begin{equation}
\begin{split}
\lefteqn{ \bra{\bm{R}',\QQ}H^0_{eh}\ket{\bm{R},\QQ} =} \\
&= \left\{ \begin{array}{cl} 
2\Delta + 3t_\perp^2/\Delta                                     & \text{if } \bm{R} =\bm{R}' \\
\frac{t_\perp^2}{2\Delta}\left(1+e^{i\QQ\cdot(\RR-\RR')}\right) & \text{if } |\bm{R}-\bm{R}'|=\tau \\
0                                                               & \text{otherwise}
\end{array}\right. \; ,
\end{split}
\label{eq:H0cv}
\end{equation}
$\tau$ being $|\bm{\tau}_\alpha|$ for any $\alpha=1,2,3$.
The Coulomb part reads 
\begin{equation}
\bra{\RR',\QQ}\,\Xi\,\ket{\RR,\QQ} = \left\{ \begin{array}{cl}
 \Xi_{\bm{R}} & \text{for } \RR = \RR' \\
 0          & \text{otherwise}
 \end{array}  \right. \, ,
 \label{eq:Coulomb}
\end{equation}
with $\Xi_\RR$ an appropriate attractive potential. In 2D, an adjustable Keldysh potential seems to be the most pertinent choice.\cite{Galvani2016} Since the effective hopping integrals now depend on the direction of the hopping step, the symmetry of the problem is much lower, but calculations can easily be done with the same techniques as for $\bm{Q}=0$. See Galvani \emph{et al.}\cite{Galvani2016} and Appendix \ref{TBdisp} for the details of the derivation.

At variance with the optical limit, at finite $\QQ$ the exchange term $V$ shall not be neglected. In fact it is responsible for dipole-like coupling between different sites which induces a singularity in the exciton dispersion at $\QQ \to 0$.\cite{Toyozawa2003} In our TB model it produces effective interactions $\bra{\RR',\QQ}V\ket{\RR,\QQ}$ when the electron and the hole are sufficiently close (within the $1nn$ shell). In the mixed representation these interactions are therefore local with respect to the electron-hole relative distance $\RR$, but have short- and long-range components with respect to the propagation of the center of mass $\QQ$. The short-range components produce analytic terms that shift upward the dispersion curve with respect to states where $V=0$, such as triplet states. Instead, the long range contribution is a dipole-dipole term whose Fourier transform is linear for small $\QQ$ and singular at $\QQ=0$, as expected. As shown in Appendix \ref{exchange}, an approximate expression valid for the single layer reads:
\begin{equation}\begin{split}
\lefteqn{\bra{\RR',\QQ}V \ket{\RR,\QQ} \approx} \\
&\approx \left\{\begin{array}{l} 0 \quad\text{for }\RR \text{ or } \RR' \notin \{\bm{\tau}_1,\bm{\tau}_2,\bm{\tau}_3\}\\
\frac{J}{4Q\tau} (e^{-i\bm{Q}\cdot \RR} - 1) (e^{i\bm{Q}\cdot\RR'} - 1) \quad \text{otherwise} \end{array} \right. \,,
\end{split}\label{exch}
\end{equation}
where $J$ will be considered here as a parameter to be fitted to \textit{ab initio} calculations, and where $Q=|\QQ|$ and $\tau = |\bm{\tau}_\alpha |$ for any $\alpha$.

\subsubsection{Bulk structures}

When passing to bulk systems, some aspects complicate the model. The first is that the corresponding Bloch states are characterized by $\kk=\kk_\perp+\hat{z}\,k_\parallel$. Moreover in the AA' and AB stacking, the basis is formed of four Bloch states because of the four atoms in the unitary cell. This problem is not encountered in the ABC stacking as long as one works in the elementary rhombohedral cell with only two atoms. The second complication is connected to the anisotropy of the layered structures that requires to distinguish between in-plane and inter-plane screening. Finally, interlayer coupling is accounted for by a hopping term $t_\parallel$ linking two neighbouring sites that are vertically aligned and a parallel $2nn$ hopping $t_{2\parallel}$ between second-nearest-neighbours of different planes, again assumed equal for $B-B$ and $N-N$ hopping. A sketch of the hopping terms in the AA' case is given in Figure~\ref{fig:triangular_lattices}(b).

Once properly generalized and retaining only $1nn$ intra- and inter-plane couplings, the energy of the $\pi^*$ conduction states reads:
\begin{eqnarray}
E_{\bm{k}e}^{AA'}  &\approx & \Delta +  \frac{(t_\perp |\gamma(\bm{k_\perp})|  \pm 2 t_\parallel \cos (k_\parallel c))^2}{2\Delta} \;, \label{AA'bands} \\
E_{\bm{k}e}^{ABC}  &\approx & \Delta +  \frac{|t_\perp \gamma(\bm{k_\perp}) + t_\parallel e^{i k_\parallel c}|^2}{2\Delta} \;, \label{rh}  \\
E_{\bm{k}e}^{AB}  &\approx & \Delta + \frac{\left(t_\parallel \cos( k_\parallel c)\pm \sqrt{(t_\parallel \cos( k_\parallel c))^{2}+t_\perp^2 |\gamma(\bm{k_\perp})|^2}\,\right)^2}{2\Delta} \;,\label{ABbands}
\end{eqnarray}
where we recall that $c$ is the interlayer distance.
For valence states $E_{\kk h}^\sigma = -E_{\kk e}^\sigma$ with $\sigma=$AA', AB, or ABC.
In the case of AA' and AB stacking the total number of bands is four (two $e$ and two $h$ bands), while the ABC stacking has only two bands consistently with the number of atoms in the respective unitary cells. However, in order to report the ABC dispersion in the smaller hexagonal cell, bands have to be folded by substituting $k_\parallel c$ in Equation~\eqref{rh} with $k_\parallel c+2\pi\lambda/3$ with $\lambda=0,\pm 1$ which results in a total of six bands.
The corresponding $2nn$ expressions are unnecessarily complicated to be reported here, but exact expressions including all terms are reported in Appendix~\ref{app:formulae}. In the successive calculations, $t_{2\parallel}$ and $t_{2\perp}$ have been included. While $t_{2\perp}$ breaks the $E_h=-E_e$ symmetry in the same way as in the monolayer, the $t_{2\parallel}$ term breaks the valence-conduction symmetry of the splitting between bands of the same character (valence or conduction). These aspects will be discussed more in detail in section IV.

The generalization of the Bethe-Salpeter-Wannier equation passes through the following steps.\begin{itemize}
\item Since we will not consider vertical dispersion, the momentum $\QQ$ is still in-plane as in the single layer.
\item The most delicate part is the indexing of the relative distances between the electron and the hole since now they can localize independently on different planes. In the AA' stacking the $\Lambda_B$ and $\Lambda_N$ sublattices are swapped from one layer to the other, and are shifted in the AB and ABC cases. As a consequence the relative distance between the two particles  belongs to a lattice that depends on the localization of the two particles. We keep track of this information following the approach introduced recently in few-layer hBN\cite{Paleari2018} consisting in (i) fixing the origin on a N site of one plane, (ii) introducing shifted triangular sublattices $\Lambda+\bm{u}$ with $\bm{u}$ a proper translation vector\footnote{In the single-layer, the origin belongs to $\Lambda=\Lambda_N$ and $\RR\in\Lambda_B=\Lambda+\bm{\tau}_j$ with $j$ any of 1, 2, or 3. We will take $j=3$ for simplicity.} (including in-plane and vertical shifts $\hat{z}c$) and (iii) generalizing $\RR$ to $\RR_{\alpha,\beta}$ where $\alpha$ and $\beta$ label the sublattice occupied respectively by the electron and the hole.
\footnote{In the cited work\cite{Paleari2018}, a detailed explanation of the splitting of excitonic states is provided showing also that surface effects lead to an energetic separation between excitons localized on the inner layers and excitons localized on the outer layers.} The example of the AA' stacking is given in Figure~\ref{fig:relative-coordinates_tb}.
\item The matrix elements of the kinetic term~\eqref{eq:H0cv} are expressed in terms of $\Delta$ and in-plane and out-of-plane hopping terms according to the exciton state $\ket	{\RR_{\alpha,\beta},\QQ}$.
\item In the Coulomb matrix elements~\eqref{eq:Coulomb} we use a standard 3D $\Xi_\RR=1/(\epsilon R)$ potential instead of the Keldysh potential. We took into account the anisotropy of the material treating $\epsilon_\parallel$ and $\epsilon_\perp$ as distinct fitting parameters.
\item In the $\QQ=0$ limit in 3D, the $\QQ=0$ singularity of the Coulomb potential is stronger than in 2D since at low $\QQ$ it varies as $1/Q^2$. This induces discontinuities at $\QQ=0$ leading to an upward shift of longitudinal modes that in bulk hBN is about 1~eV.\cite{Toyozawa2003,Koskelo2017} Since the neglect of this effect does not change the conclusions regarding the nature of the gapwidth, we neglected the exchange term~\eqref{exch}.
%for the longitudinal exciton modes which moves upwards, and produces a longitudinal-transverse splitting\cite{Toyonzawa2003} as shown recently in the case of AA' hBN by Koskelo and co-workers.\cite{Koskelo2017}. Since these longitudinal modes are not optically active, we will not take into account these long range exchange contributions in the following.

%exchange $V$ is not linear in $\QQ$ but it diverges as $\sim 1/Q^2$. This singularity splits the transverse exciton and the longitudinal one,\cite{Toyozawa2003} moving the latter to energies as high as 1 eV in the case of hBN.\cite{Koskelo2017}. Since longitudinal modes are not optically active, and in this case they are well separated from the lowest energy excitons, we decided non to take into account these long-range exchange contributions in the following.
\end{itemize}

\section{The free-standing single layer}

\subsubsection{Single-particle band structure}
\label{singlelayer}

\begin{figure}
\centering
\includegraphics[width=0.49\textwidth]{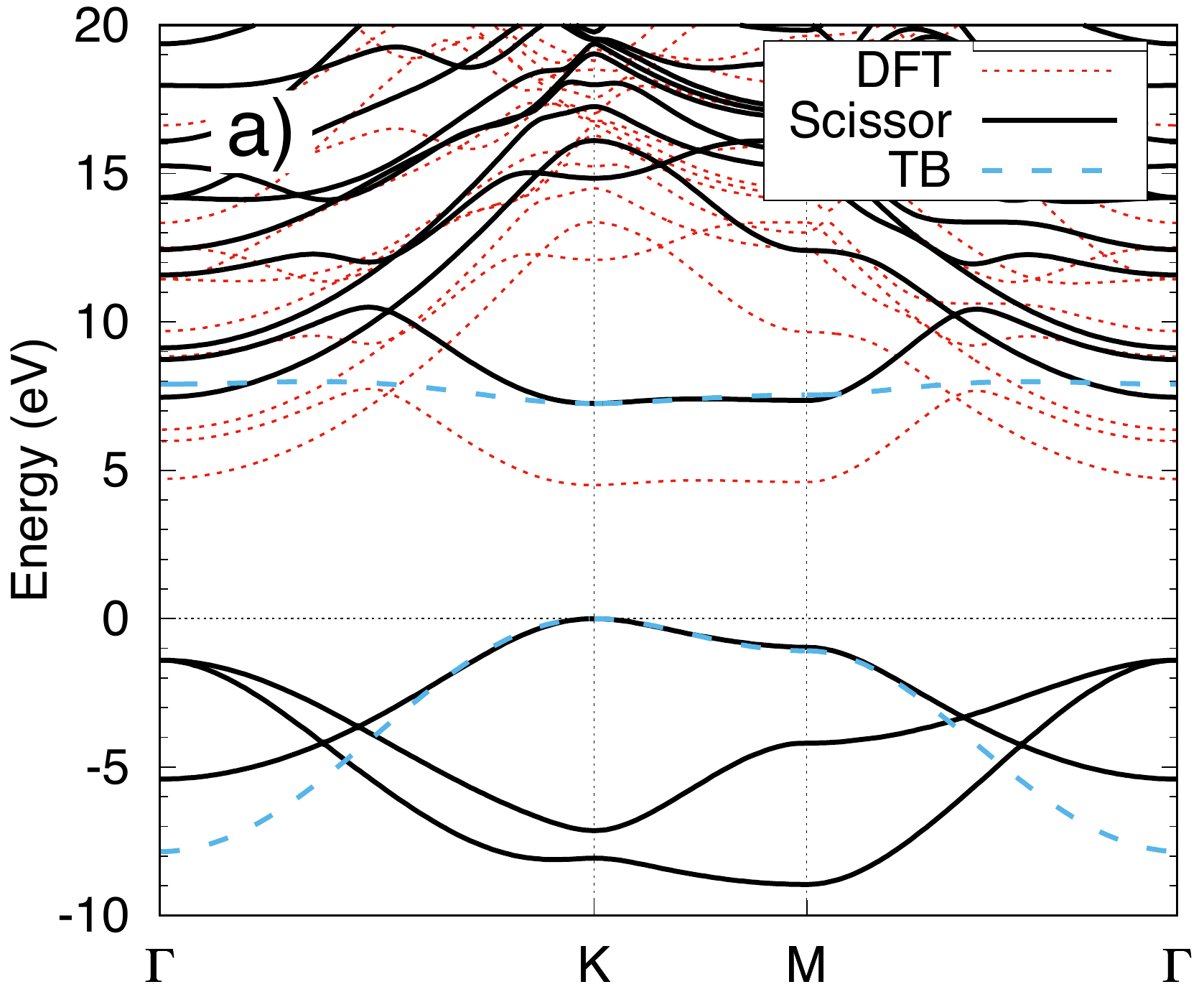}

\vspace{5mm}

\includegraphics[width=0.49\textwidth]{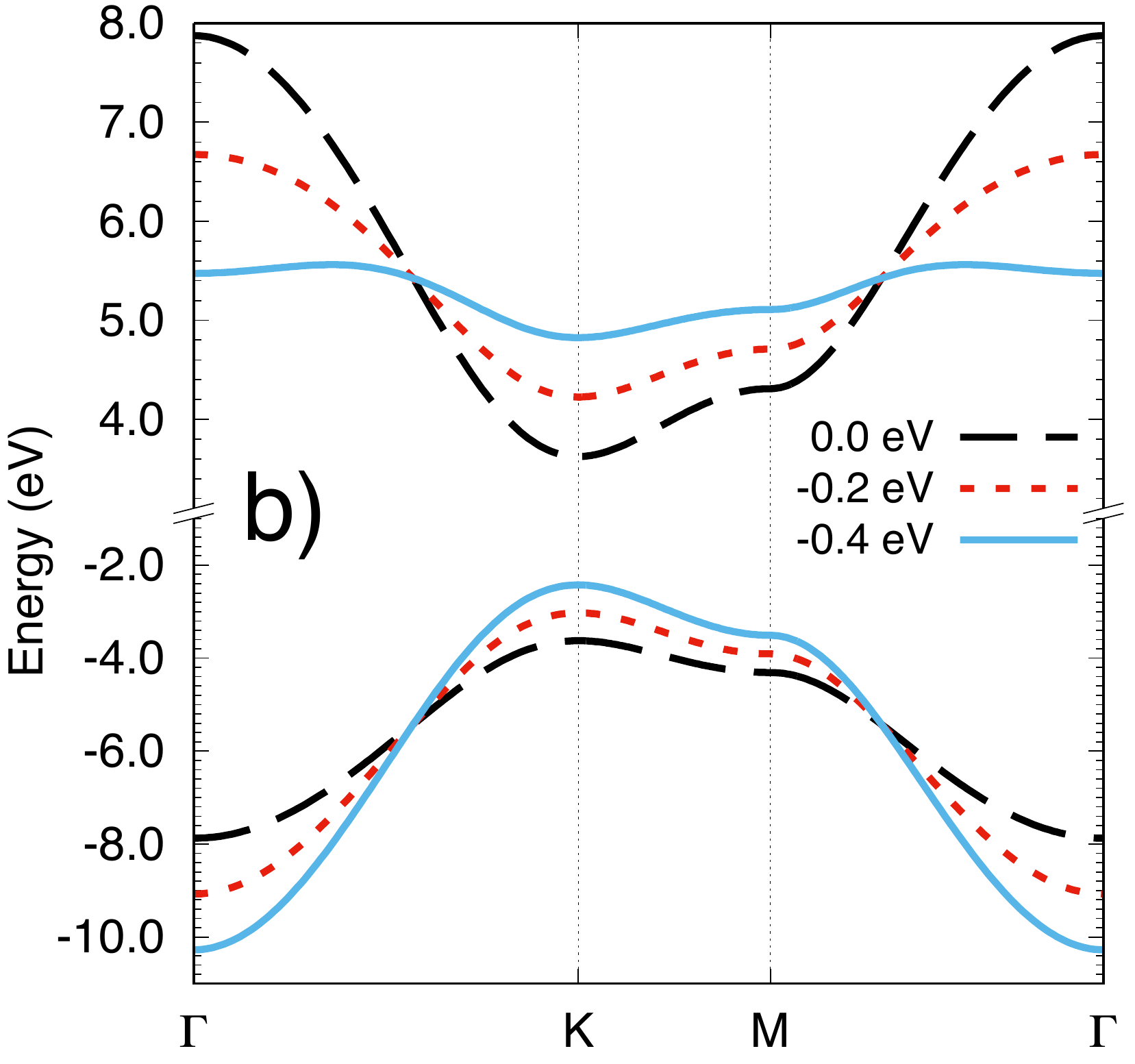}
\caption{The single-particle band structure of free-standing $h$--BN single layer (a) \emph{ab initio} DFT (dotted red), quasiparticle (solid black) calculation, and optimal TB fitting (dashed blue).
%band structures of a single hBN layer: in the MK region, the gap is direct at point K between the flat $\pi$ bands. 
(b) Dependence of the TB band structure on the parameter $t_{2\perp}$.}
\label{fig:ml1_bands}
\end{figure}

\begin{figure*}
\centering
\includegraphics[width=0.83\textwidth]{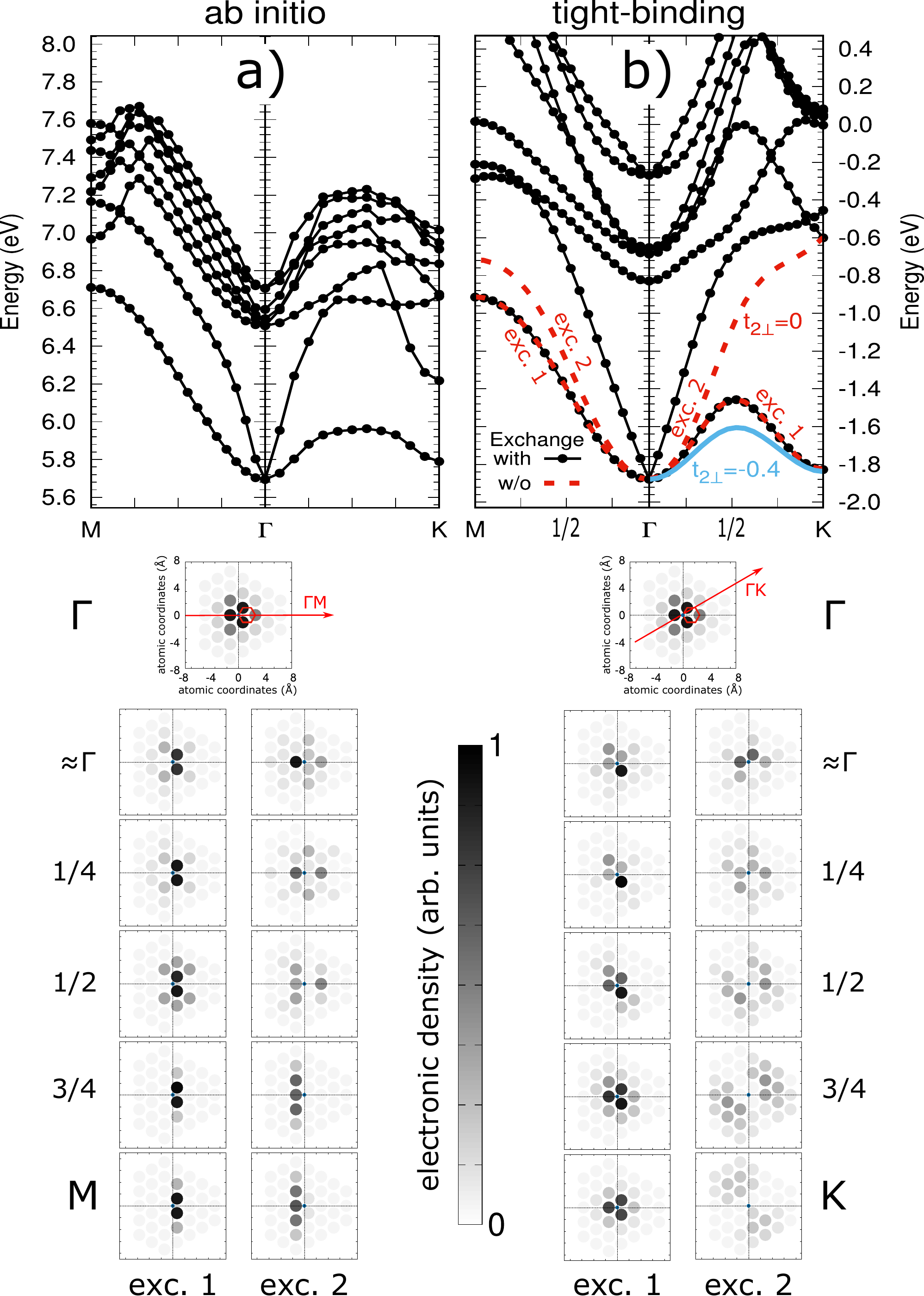}
\caption{The exciton dispersion for $\bm{q} \parallel \Gamma M$  and $\bm{q} \parallel \Gamma K$ computed with (a) \textit{ab initio} and (b) tight-binding methods. In the (b) panel, results without $2nn$ hopping and including the exchange term are reported in black, calculations without the exchange are reported in dashed-red ($t_{2\perp}=0$~eV) and solid blue ($t_{2\perp}=-0.4$~eV).\\
Bottom panels: Real-space representation of the tight-binding electronic density $|\Phi_{\RR\QQ}|^2$ of the first two excitons with the hole located at the center of the axis (blue dot). At $\Gamma$, we report in dashed-red the direction parallel to the $\QQ$ vector of the series and a red hexagon marks the real-space hexagonal cell.}
\label{fig:ml1_exc_disp}
\end{figure*}

In  Fig.~\ref{fig:ml1_bands}a we report both the DFT and the quasiparticle results for the hBN monolayer band structure. The quasiparticle gap, equal to 7.25 eV\cite{Galvani2016}, is direct between the $\pi$ and $\pi^*$ bands at point $K$ in the Brillouin zone, while the bands are very flat along the $KM$ lines. These valence and conduction bands are not completely symmetric, the dispersion being larger for the valence band. As discussed elsewhere, these results are in agreement with several previous calculations.\cite{Blase1995,Wirtz2009,Ribeiro2011,Berseneva2013,Huser2013,Cudazzo2016,Kang2016,Galvani2016}

The TB model introduced previously gives the best fit to the KM region of the \emph{ab initio} band structure with  $\Delta = 3.625 $ eV, $t_\perp= -2.33 $ eV, and $t_{2\perp}=-0.4$ eV. The fit to ab-initio data and the results presented here have not been obtained from the approximate expression~\eqref{eq:TB_single-layer_1particle}, but from the exact diagonalization of the full TB Hamiltonian reported in Appendix~\ref{app:formulae}.

The inclusion of $t_{2\perp}$ breaks the electron-hole symmetry, as clearly shown in equation~\eqref{eq:TB_single-layer_1particle}, by reducing the effective hopping integral of the conduction band and increasing that of the valence band. As a consequence the conduction band is flatter than the valence band, in agreement with the \textit{ab initio} results. The sensitivity of the band structure to the $2nn$ contribution is exemplified in Fig.~\ref{fig:ml1_bands}b where different values of $t_{2\perp}$ have been used. Note that the (direct) gap at $K$ is not modified because the symmetry of the crystal leads to $\gamma(\mathbf{k_\perp}=K)=0$.
To conclude, this simple TB model is able to reproduce the $\pi$ states in the regions where excitons are relevant for the optical properties.

\subsubsection{Exciton dispersion}

The dispersion curves $E_{exc}(\bm{Q})$ obtained from \textit{ab initio} calculations are shown in Fig.~\ref{fig:ml1_exc_disp}a. When $\bm{Q}=\Gamma$ we recognize the excitons already characterized in previous works, with the doubly degenerate ground state exciton of symmetry $E$ (or $1s$ in the atomic-like notation).\cite{Galvani2016,Koskelo2017} At higher energy we have the dispersion of the six $2s$ and $2p$ states. Note that in either direction one can recognize some additional parabolic bands at high energy close to the zone boundary. We will come back to this characteristic later, when discussing the TB model.
In both $\Gamma M$ and $\Gamma K$ directions, the second exciton is much more dispersing than the ground-state one. Actually it has been shown that its linear dispersion at $\QQ\to 0$ is a peculiarity of the 2D geometry which is generated by the exchange contribution to the electron-hole interaction kernel in the Bethe-Salpeter equation.\cite{Cudazzo2016,Qiu2015} Finally we highlight the weak dispersion of the first exciton along $\Gamma K$ and in particular the fact that the energy at $\QQ=\Gamma$ and $\QQ=K$ basically coincide. This is expected because the most intense single-particle transitions (and thus the most important contributions in the excitonic spectrum) at $\QQ=\Gamma$ and $\QQ=K$ come from vertical $K\to K$ and slant $K\to K'$ transitions respectively. Since the single-particle states $K$ and $K'$ have the same energy, the resulting dispersion of the exciton along $\Gamma K$ is expected to attain the same value at the extrema of the path. Our results along $\Gamma M$ are very similar to those published in literature.\cite{Cudazzo2016,Koskelo2017}

The exciton dispersion has been also computed within the TB approach and is reported in Figure~\ref{fig:ml1_exc_disp}b. The $J$ term of the exchange contribution~\eqref{exch} has been treated as a fitting parameter, fixed here at 5 eV. The curves have been obtained by diagonalizing a matrix involving 860 sites with a Keldysh potential ranging up to the 9th shell.
At moderate values of $Q$ the TB model agrees very well with \textit{ab initio} results. In particular it reproduces the double degeneracy of the ground state at $\bm{Q}=0$. The two corresponding states $\ket{\Phi^+}$ and $\ket{\Phi^-}$ can be taken as two ``circular'' states whose components on the three $\bm{\tau}_{1,2,3}$ sites $\braket{\bm{\tau}}{\Phi^{+}}$ and $\braket{\bm{\tau}}{\Phi^{-}}$ are proportional to the cubic roots of unity $(1,\omega,\omega^2)$ and $(1,\omega^2,\omega)$, respectively.

As soon as one moves away from $\QQ=0$ the degeneracy is lifted according to their different dipolar orientation. The lower-energy exciton has a transverse orientation at low $\QQ$, i.e. the electron-hole dipole is perpendicular to the momentum $\QQ$. This makes it optically active and its dispersion is insensitive to the exchange term. Instead, the higher energy exciton is longitudinal, its dipole being parallel to $\QQ$. This makes it optically dark and particularly sensitive to the exchange term which has a linear dependence at low $\QQ$. Indeed Figure~\ref{fig:ml1_exc_disp}(b) shows that the degeneracy between the two excitons is mildly lifted by the Coulomb term alone (red dashed line), whereas the inclusion of the exchange term reproduces the correct linear dispersion.
In Appendix~\ref{app:perturb} we report an analytical result predicting this behaviour within a perturbative treatment of the exchange interaction.\cite{Wu2015,Qiu2015}

At large $Q$, the agreement is less satisfying since the formula used is no longer sufficient (terms involving sums over reciprocal lattice vectors are neglected) and also because only $\pi$ states are considered. This consideration allows us to point out the origin of the parabolic bands observed in the \textit{ab initio} calculation close to the zone boundary. These are present also in the TB model along $\Gamma K$, so we can ascribe these bands to $\pi \to \pi^*$ excitations. Instead, along $\Gamma M$ these bands are predicted only in the \textit{ab initio} solution, so we can advance the hypothesis that they are of the $\sigma \to \pi^*$ or $\pi \to \sigma^*$ type as the $\sigma$ states are absent in our TB model.

Further improvements on the TB calculation are obtained when $2nn$ hopping integrals are included, as a consequence of the fact that they break the valence-conduction symmetry of the band structure. When $t_{2\perp}=-0.4$ eV, the main effect is to decrease the dispersion of the lowest exciton along $\Gamma K$  from 0.5~eV to 0.3~eV, close to the \textit{ab initio} value of about 0.27 eV, as reported in Figure~\ref{fig:ml1_exc_disp}b). In view of the simplicity of the TB model the overall agreement is very good. 

To go beyond, we can  represent the excitonic wave function as a function of $\bm{R}$, which is the distance between the electron and the origin where the hole is fixed.  Indeed, in the mixed $(\RR,\bm{Q})$ representation and for fixed $\bm{Q}$, the excitonic wave function can be expanded in the corresponding space:
\begin{equation}
\ket{\Phi} =   \sum_{\bm{R}}\Phi_{\bm{RQ}} \ket{\bm{R},\QQ} \; .
\end{equation}
As discussed in Appendix \ref{TBdisp}, the basis $\ket{\bm{R},\QQ}$ and therefore the coefficients $\Phi_{\bm{RQ}}$ are not uniquely defined. With our standard definition $(\bm{k_e},\bm{k}_h) = (\kk,\bm{k}-\bm{Q})$,  $\Phi_{\bm{RQ}}$ is in general a complex quantity. In the bottom part of Figure~\ref{fig:ml1_exc_disp} we show a map of the intensities $|\Phi_{\bm{RQ}}|^2$ (electronic densities) which are gauge invariant. Here we consider more particularly the first two excitons (degenerate at $\bm{Q}=0$) along $\Gamma M$ and $\Gamma K$.

Consider first the $\Gamma M$ direction (left panels). Since this is a mirror of the point group we expect the wave function to be either odd or even with respect to the reflection symmetry so that the dipole $\bm{d}_\Phi$ is either perpendicular to $\QQ$ (transverse mode, exciton 1) or parallel to it (longitudinal mode, exciton 2). In particular the intensity corresponding to the transverse mode should vanish on the symmetry axis. As shown in the plots, this is clearly the case for exciton 1 starting from the degenerate ground state exciton at $\bm{Q} =0$. Actually both excitons remain fairly localized, but they deform significantly as a function of $\bm{Q}$ with a tendency to become elongated in a direction normal to $\Gamma M$. In the $\Gamma K$ case (right panels), we no longer expect definite symmetries except at $\Gamma$ and $K$ points, but we can see in the figure that the first exciton remains fairly localized with a compact shape except at the middle of $\Gamma K$ where it tends to elongate. At low $\bm{Q}$  the system can be considered as quasi-isotropic, but still we have longitudinal and transverse modes.  

Let us finally mention an interesting limit for the ground state exciton in the extremely localized case, where the excitonic wave function only extends to the $1nn$ of the (fixed) hole. It turns out that this model can be completely solved in real space. The resulting lowest mode does not disperse at all in the whole Brillouin zone. This type of flat band has attracted recently great interest in various fields of solid state physics.\cite{Zheng2014} The corresponding toy excitonic model is described in Appendix \ref{kagome}.

\section{The AA', AB, and ABC stacking} 

It is an established theoretical result that the electronic gap changes from direct at $K$ in the monolayer, to indirect in the AA' stacking.\cite{Arnaud2006,Arnaud2008,Wirtz2008,Galambosi2011,Sponza2018} This effect has important consequences in the optical properties of bulk hBN, as it has been stressed by some recent works.\cite{Cassabois2016a,Cassabois2016b,Schuster2018,Schue2018,Sponza2018} Motivated by this, we investigate the impact of the stacking sequence on the single-particle (band structure, electronic gap) and two-particle excitations (exciton dispersion, optical gap).

\subsection{Single-particle band structure}

\begin{figure}
\centering
\includegraphics[width=8cm]{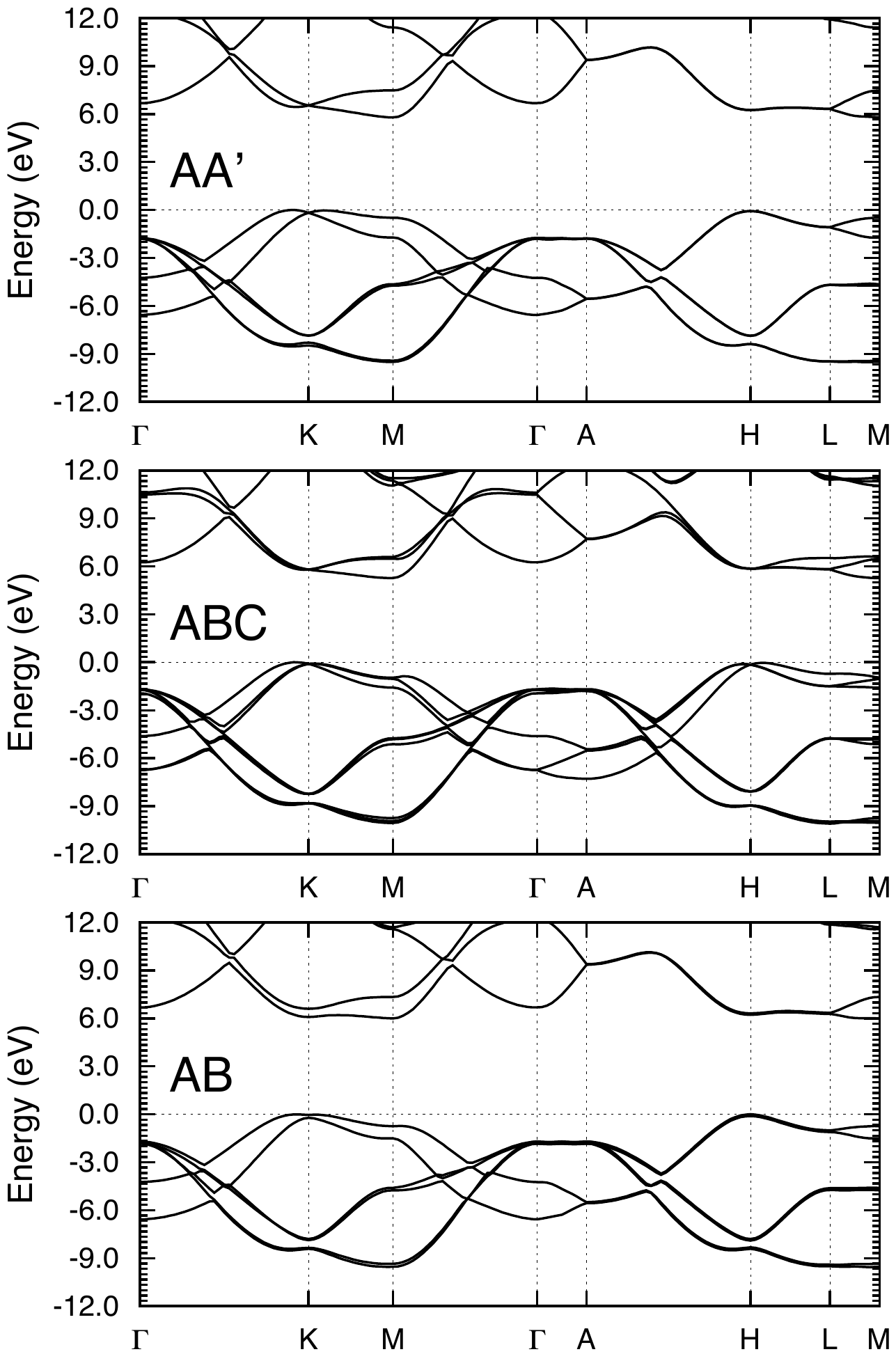}

\vspace{5mm}
\includegraphics[width=0.5\textwidth]{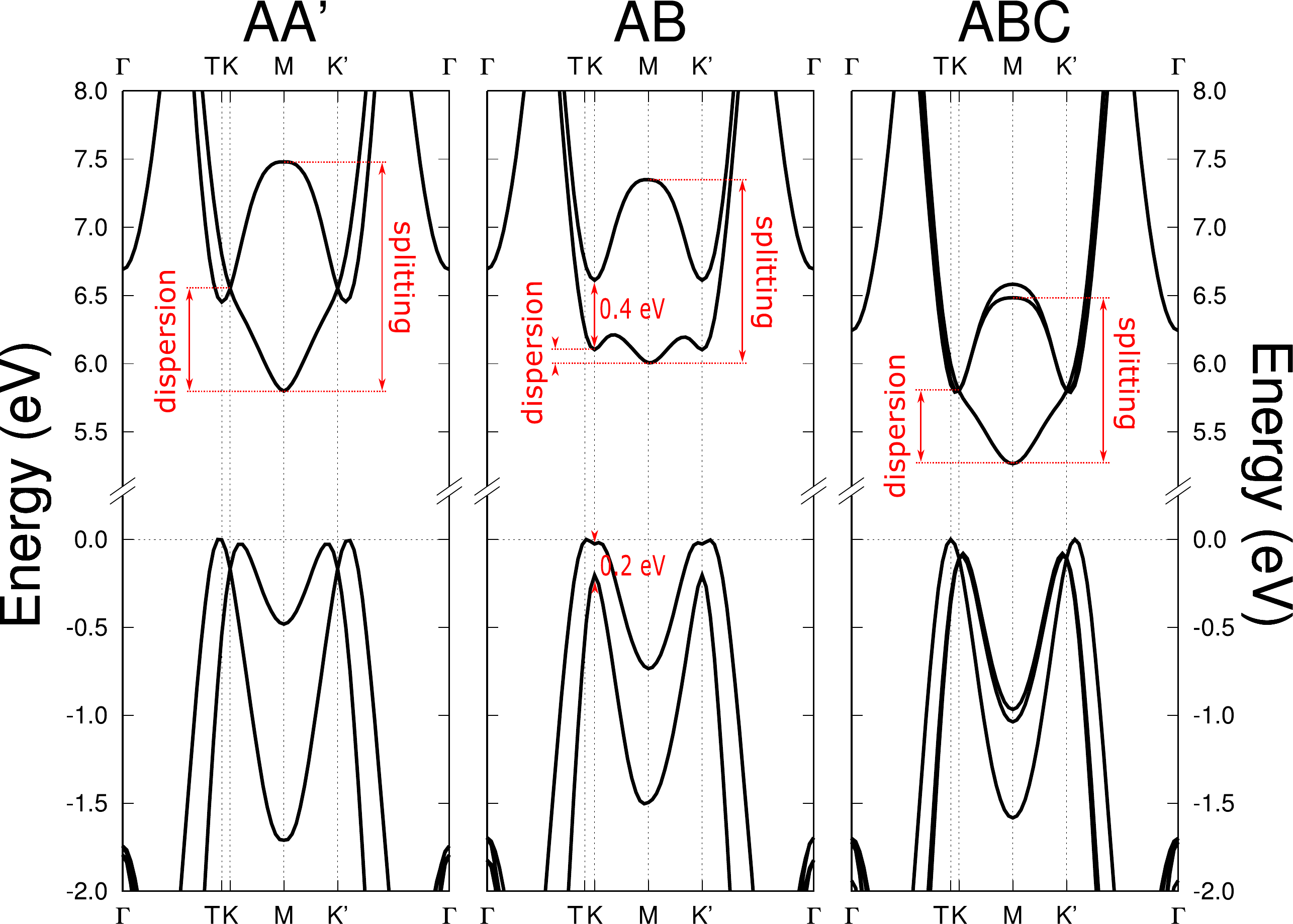}
\caption{The GW band structure of the three bulk phases drawn in the hexagonal Brillouin zone.}
\label{fig:bulk_bands-full}
\end{figure}

\begin{table}
\centering
\begin{tabular}{lcc} 
\hline \hline
stacking & dispersion & splitting\\
\hline 
AA'  & 0.75 & 1.68 \\
AB   & 0.10 & 1.34 \\
ABC & 0.51 & 1.11 \\
\hline \hline
\end{tabular}
\caption{Dispersion along $KM$ and splitting at $M$ of the GW bottom conduction (see bottom panels of Fig.~\ref{fig:bulk_bands-full}). 
All energies are in eV. }
\label{tab:bands_stacking}
\end{table}

The GW full band structure of the three stackings is reported in Fig.~\ref{fig:bulk_bands-full} along high-symmetry lines of the hexagonal Brillouin (for the band folding of the rhombohedral ABC structure, see explanation in section~\ref{sec:methods:ab-initio:bulk}) while the relevant quasiparticle dispersion along the $K M K'$ path is plotted in the bottom panels of Figure~\ref{fig:bulk_bands-full} where its principal characteristics (dispersion along $\Gamma M$ and splitting at $M$) are highlighted in red. Results are also summarized in Table~\ref{tab:bands_stacking}. The first interesting feature is the behaviour of the highest valence and the lowest conduction bands at the $K$ point. In the monolayer, the valence and conduction bands have their extrema at  $K$, but this is not the case in the AA' stacking. Here two valence bands cross each other yielding two local maxima in two points close to $K$. We indicate with letter $T$ the one along $\Gamma K$. In the conduction region, two bands also cross at $K$ but they form only one local minimum at $T$.  In the ABC stacking, the $K$ point in the  rhombohedral Brillouin zone has no particular symmetry,\cite{Ho2016} and even if the folded bands show extrema there, they do not correspond to global extrema of conduction and valence bands.

Qualitatively very different is the AB stacking, where the crossing is avoided in both valence and conduction bands because of symmetry reasons, leading to a pretty flat dispersion of the top valence in the vicinity of $K$. In the conduction band, the same avoided crossing yields a clear local minimum at $K$. The splitting at $K$ between the two highest occupied states is about 0.2~eV, and it is about 0.4~eV between the two lowest empty states.

A second interesting aspect is the peculiar dispersion of the lowest conduction band along the $KM$ direction, which is the most relevant direction for the optical properties of this material.\cite{Schuster2018,Sponza2018,Schue2018} The qualitative behaviours of the AA' and the ABC phases are similar: away from $K$ the lowest conduction band disperses almost linearly and has a minimum at $M$, while in the case of the AB stacking the dispersion is flatter and has a concave shape away from $K$. Still, beyond a local maximum between $K$ and $M$ it also attains its minimum at $M$. The dispersion of the bottom conduction is 0.75~eV in the AA', 0.51~eV in the ABC and only 0.10~eV in the AB.
It is also worth reporting the energy splitting between the two lowest conduction bands at $M$: this is 1.68~eV in the AA' stacking, 1.11~eV in the ABC stacking, and 1.34~eV in the AB stacking. These data are reported also in Table~\ref{tab:bands_stacking}.

The nature of the gap also merits to be discussed. The smallest direct and indirect gaps extracted from the band structure are reported in Table~\ref{tab:gap_stacking} for the GW and the LDA band structure. One immediately sees that the three structures have similar gaps at the LDA level, and in particular there is negligible difference between the AA' and the ABC structures. Instead, after the inclusion of GW corrections, the gaps (direct and indirect) of the ABC stacking are sensibly smaller than those of the other two structures, mostly because of the different quasiparticle corrections to the valence bands.
% and are due for a large part to dynamical effects of the correlation self-energy.
%However, self-consistent approaches such as the quasiparticle self-consistency GW or the energy self-consistent GW$_0$ may change this perturbative result. These analyses though go beyond the scope of our work since the very dispersion of the band is quite insensitive to the lack of self-consistency.
This result suggests that there are significant differences in the screening properties of the ABC stacking with respect to those of the other two phases, but a detailed analysis in this respect goes beyond the scope of this article.
Finally, note that  in the AA' phase, the smallest direct gap is at $M$. However in $T$, where optical matrix elements are stronger, it is 4.64 eV and 6.45 eV in LDA and GW respectively. In the AB stacking it is located at $K$ and in the ABC stacking actually does not lie on a high-symmetry line of the hexagonal cell. In the latter case, the smallest direct gap is actually 5.75~eV, so very close anyway to the direct gap at $K$ reported in Table~\ref{tab:gap_stacking}.

\begin{figure}
\begin{center}
\includegraphics[width=1.\linewidth]{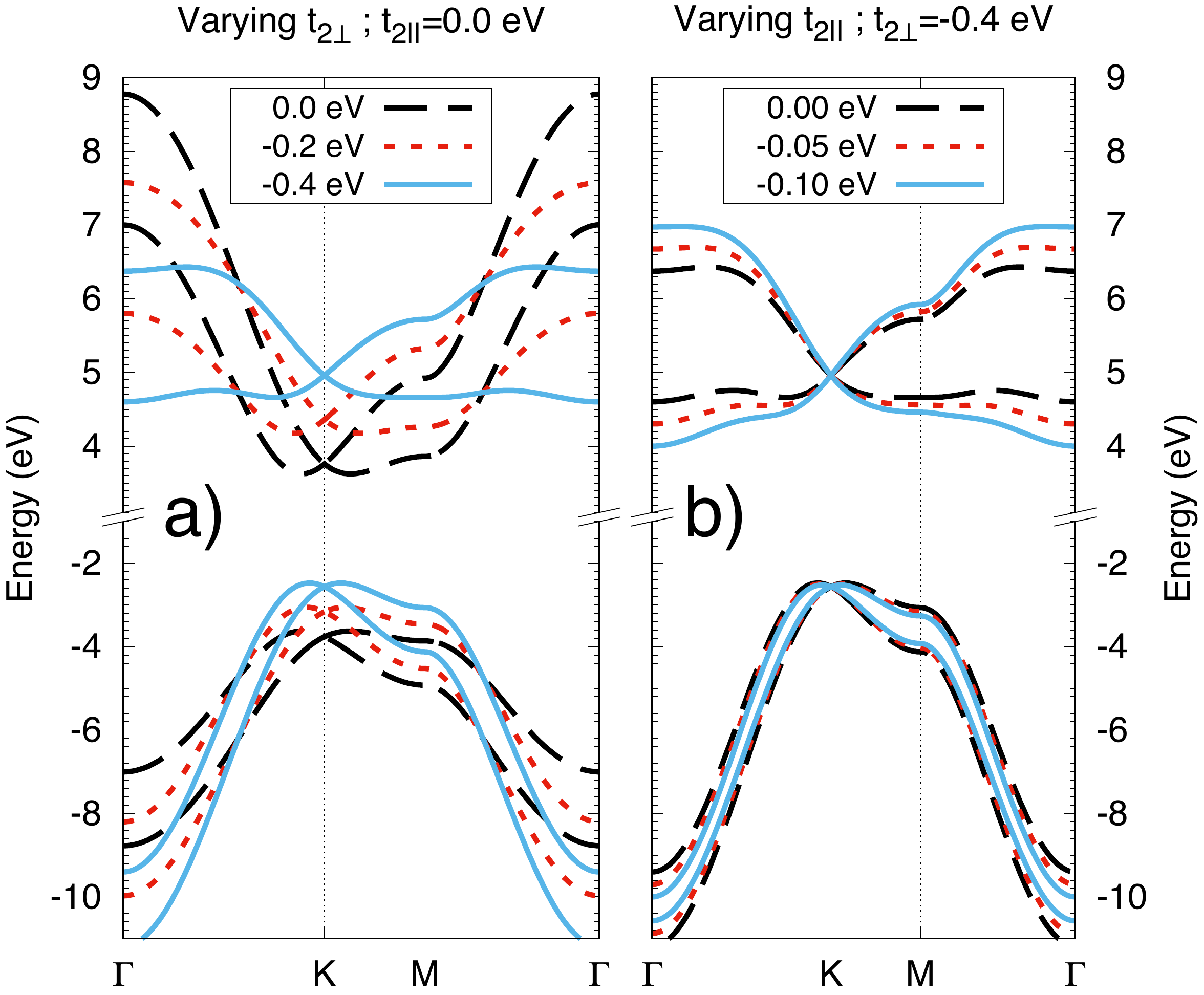}
\caption{Electronic band structure of the AA' phase. Influence of the TB $2nn$ interactions: (a) in-plane $t_{2\perp}$ and (b) between planes $t_{2\parallel}$.}
\label{TBbands}
\end{center}
\end{figure}

\begin{table}
\centering
\begin{tabular}{l| ccc | ccc }
\hline \hline
  & \multicolumn{3}{ |c| }{Smallest direct} & \multicolumn{3}{ c }{Indirect} \\
  & GW & LDA & point & GW & LDA & points \\
\hline 
AA' &  6.28 & 4.46 & $M$ & 5.80 & 4.02 & $TM$ \\
AB  &  6.13 & 4.31 & $K$ & 6.01 & 4.20 & $TM$ \\
ABC &  5.82 & 4.51 & $K$ & 5.27 & 3.95 & $KM$ \\
\hline
\end{tabular}
\caption{The smallest direct and indirect gaps (GW and LDA) extracted from the bandplot of the three structures.
Energies are in eV.}
\label{tab:gap_stacking}
\end{table}

Consider now the $AA'$ stacking treated within TB. In the simplest approximation we keep the same $\Delta$ and $t_\perp$ as in the monolayer,\footnote{In principle all parameters should be adjusted; in particular the gap equal to $2\Delta$ should be reduced, but as far as orders of magnitude are concerned this will not modify our discussion.} and we add only the first-neighbour interlayer hopping $t_\parallel$. Then the two conduction $\pi^*$ eigenvalues can be approximated by Equation~\eqref{AA'bands} (and with opposite sign the two valence $\pi$ states). From the equation we get that the splitting between the $\pi$ states vanish either when $\gamma(\bm{k}_\perp) = 0$ ($HK$ line, not shown here) or when $k_\parallel c =\pm \pi/2$, \textit{i.e.} on the upper and lower faces of the Brillouin zone ($AHL$ line). In fact since the periodicity along $z$ is $2c$, then $k_\parallel c \in [-\frac{\pi}{2},\frac{\pi}{2}]$. Actually the doubly degenerate state dispersing along $HL$ has exactly the average energy of the two splitted branches on $\Gamma KM$. The model reproduces the bottom conduction at $M$, where the splitting between the two conduction bands is $4t_\perp t_\parallel/\Delta$ ($|\gamma(M)|=1$). The fitting procedure to the \emph{ab initio} band structure yields $t_\parallel \simeq 0.5$ eV,\textit{ i.e.} $t_\parallel/t_\perp \simeq 0.2$ which indicates that the interlayer coupling is actually fairly important.

Sticking to  the first-neighbour level, we can already predict from Equations.~\eqref{rh} and \eqref{ABbands} that the conduction-band splitting at $M$ will decrease along the series AA', AB, ABC, and similarly for the valence-band splitting.  Indeed  in the AB stacking the splitting $2t_\parallel \sqrt{t_\parallel^2+t_\perp^2}/\Delta$ is smaller than in the AA' configuration because the number of hetero-atomic pairs along the stacking axis is smaller on average. Even smaller it is in the ABC stacking, where the splitting $\frac{3}{2} t_\perp t_\parallel / \Delta $ is indeed the lowest.\footnote{Similar analyses have been performed in the case of carbon layer stackings.\cite{Charlier1994}} Moreover in the case of the AB stacking we can verify from Eq.(\ref{ABbands}) that the bands do not cross at $K$.

However,  to reproduce the electron-hole asymmetry between the conduction and valence bands, and hence the presence of an indirect gap,  we need at least second-neighbour interactions within the planes ($t_{2\perp}$) and between the planes ($t_{2\parallel}$), as discussed in section \ref{singlelayer}. The latter term accounts for the difference of splitting, larger in the conduction band than in the valence band by about 0.4 eV in the AA' phase. Approximated formulae are reported in Appendix~\ref{app:formulae} together with a summary of the parameters used. In Figure~\ref{TBbands} we report a study of the changes induced in the AA' band structure by variations of the $2nn$ hopping terms. Typical orders of magnitude are $t_{2\perp} \simeq -0.4$ eV and $t_{2\parallel} \simeq -0.1$ eV. Using similar parameters the band structures of the other stackings are also well reproduced within the TB approximation.

\subsection{Exciton dispersion}

Let us now pass to the discussion of how the stacking sequence, and hence the changes in the band structure, affect the exciton dispersion.
Preliminary results for the AA' phase can be found in some previous works of ours.\cite{Schue2018,Sponza2018}
In Figure~\ref{fig:excitons_stacking} we report \textit{ab initio} calculations of the exciton dispersion (black curves) and the free-carrier dispersion (red curves) in the three bulk phases.
Quantities related to these dispersion relations and the exciton binding energy in the three systems are also reported in Table~\ref{tab:excitons_stacking}.

\subsubsection{Ab initio calculations}

\begin{figure}
\centering
\includegraphics[width=0.43\textwidth]{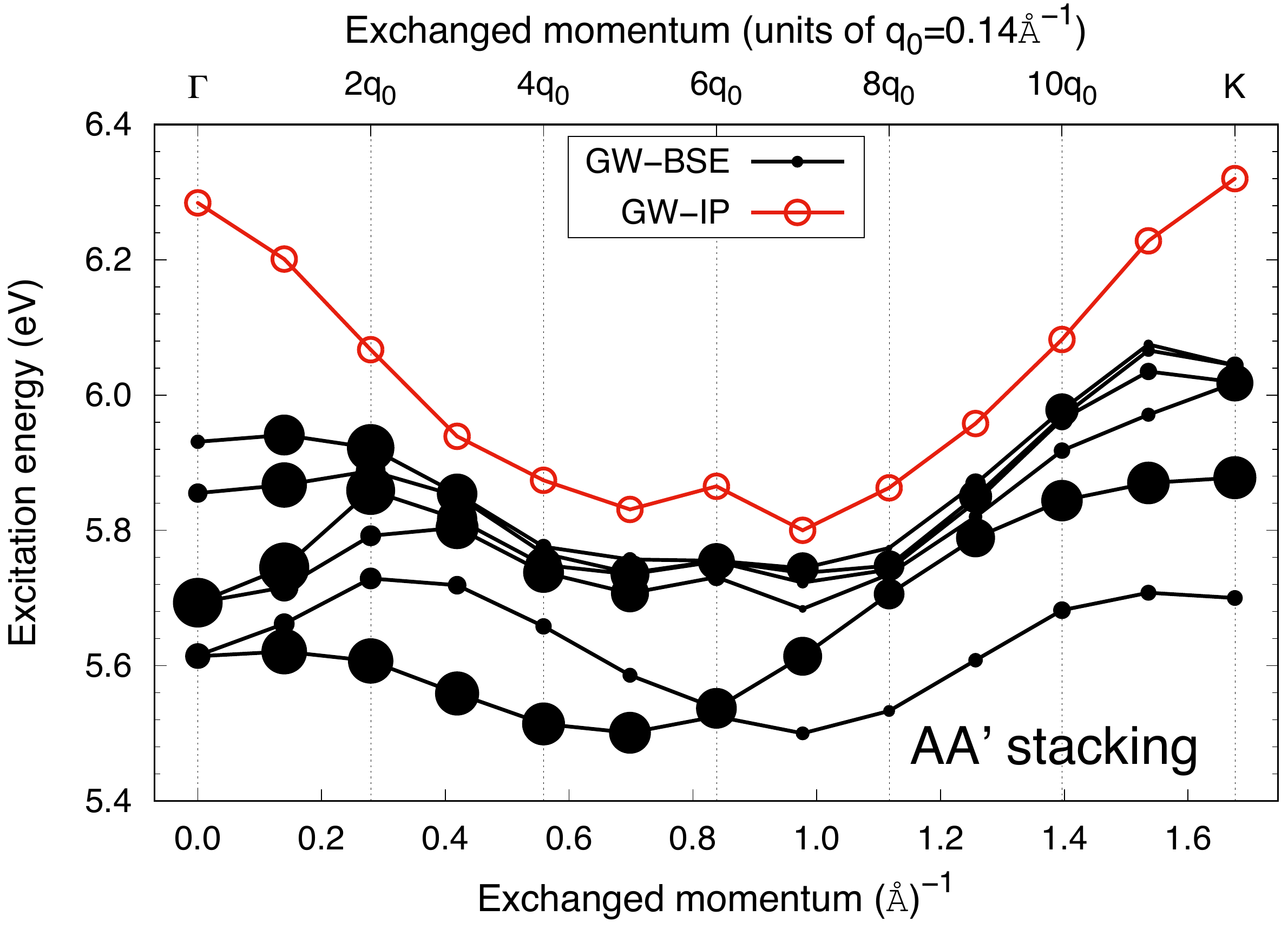}\\
\vspace{5mm}
\includegraphics[width=0.43\textwidth]{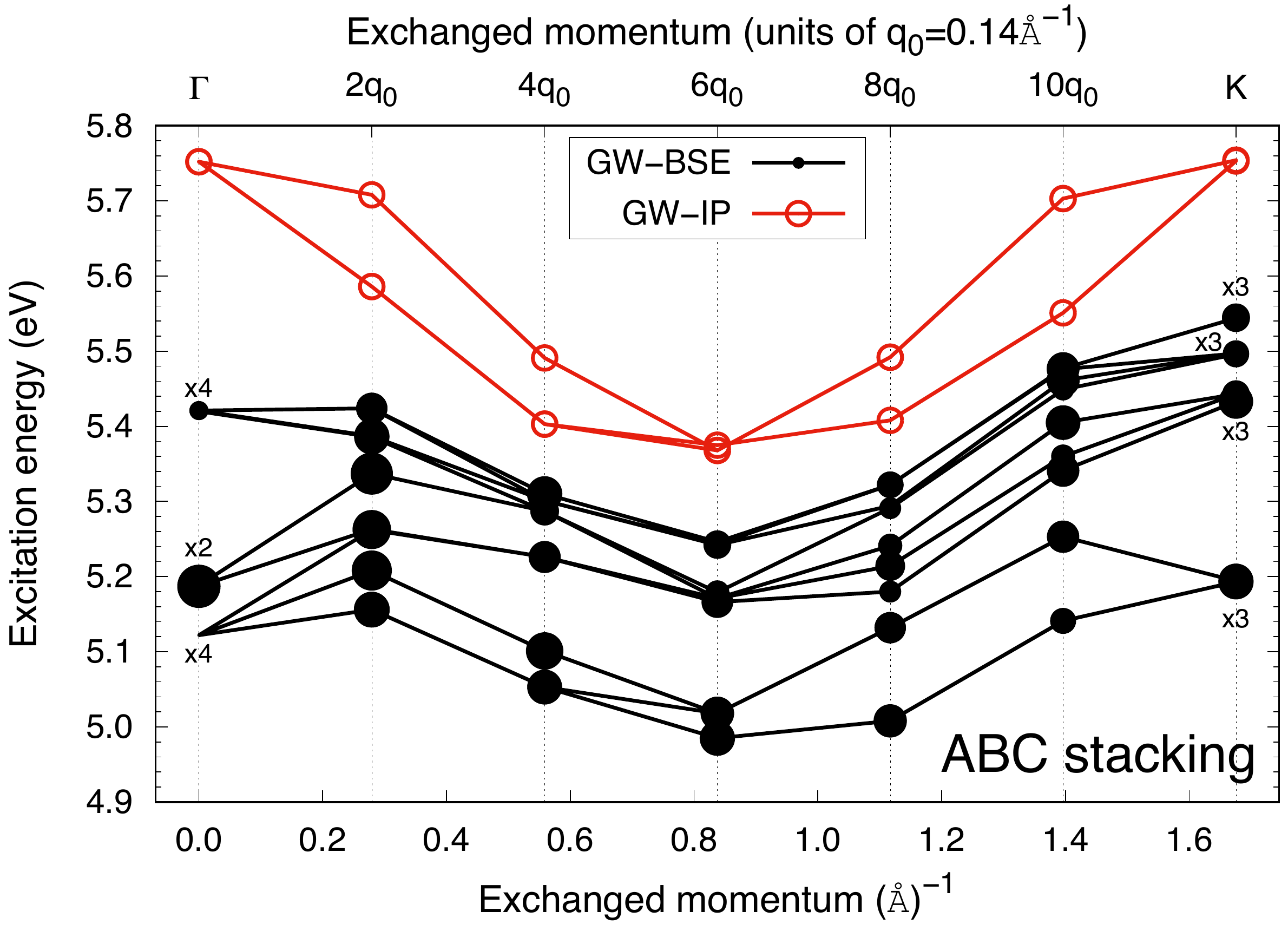}\\
\vspace{5mm}
\includegraphics[width=0.43\textwidth]{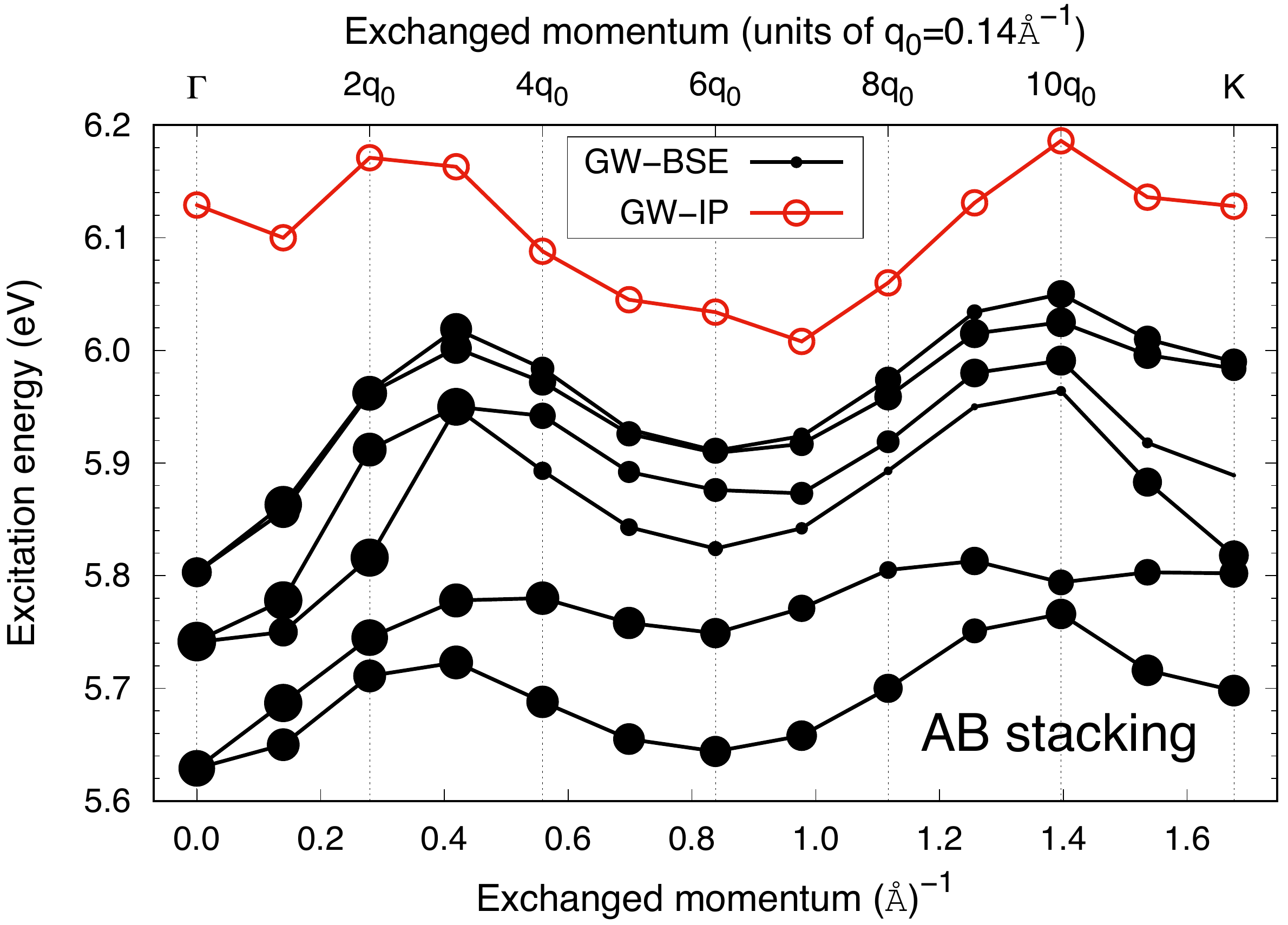}
\caption{From top to bottom: Dispersion of the first excitons in the AA', ABC and AB stacking for $\mathbf{Q}\parallel \Gamma K$  (black bullets: GW-BSE). The size of the spots is proportional to the logarithm of the intensity in arbitrary units.
The dispersion of the lowest independent-particle transition (free-carrier dispersion) is reported with red empty circles (GW-IP).}
\label{fig:excitons_stacking}
\end{figure}

\begin{table}
\centering
\begin{tabular}{lccc}
\hline \hline
       &  AA'  &  ABC &  AB   \\
\hline
GW-IP  &  0.48 & 0.38 &  0.12 \\
GW-BSE &  0.11 & 0.14 & -0.02 \\
Min. of $E_b$ & 0.30 & 0.35 & 0.35\\
Max. of $E_b$ & 0.67 & 0.63 & 0.50\\
\hline \hline 
\end{tabular}
\caption{Dispersion of the first free-carrier transition (GW-IP) and first exciton (GW-BSE). Minimum and maximum of the binding energy $E_b(\QQ)$. All energies are in eV.}
\label{tab:excitons_stacking}
\end{table}

First, let us focus on the independent-particle dispersion, or the free-carrier dispersion (GW-IP curves). In the AA' case, one recognises in the convex shape with a minimum at the middle of the $\Gamma K$ distance the dispersion of the bottom conduction along $KM$, with a peculiar double-dip shape reminiscent of the conduction band at $T$.  All along the GW-IP dispersion, the energy of the transition passes from 6.28~eV at $\mathbf{Q}=\Gamma$, corresponding to the smallest direct gap at $M$, to a minimum of 5.80~eV at 1~\AA$^{-1}$, corresponding to the indirect gap $TM$, as already reported in Table~\ref{tab:gap_stacking}. The resulting dispersion is 0.48~eV. Qualitatively, the same shape characterizes also the dispersion of the first exciton (GW-BSE), but the electron-hole kernel of the Bethe-Salpeter equation has the effect of enhancing the localization of the electron close to the hole and hence of quenching the exciton dispersion to only 0.11~eV. In fact the exciton binding energy $E_b(\QQ)$ exhibits a strong dependence on the exchanged momentum, varying by almost 0.4~eV throughout the $\Gamma K$ path. The principal consequence is that the difference between direct and indirect exciton is smaller than what an analysis based uniquely on the band structure would suggest. As pointed out by some of us,~\cite{Schue2018}  this has an important implication in the difference between photoluminescence and absorption spectra, the former being more sensitive to the lowest (possibly indirect) excitation while the latter displays highest intensity for direct excitations.

Note that the dispersion of the first two excitons cross at the middle of the $\Gamma K$ path where they have very close energies. As the size of the dots suggests, one of them is bright (large dots) and the other dark (tiny dots).

\begin{figure}
\centering
\includegraphics[width=0.50\textwidth]{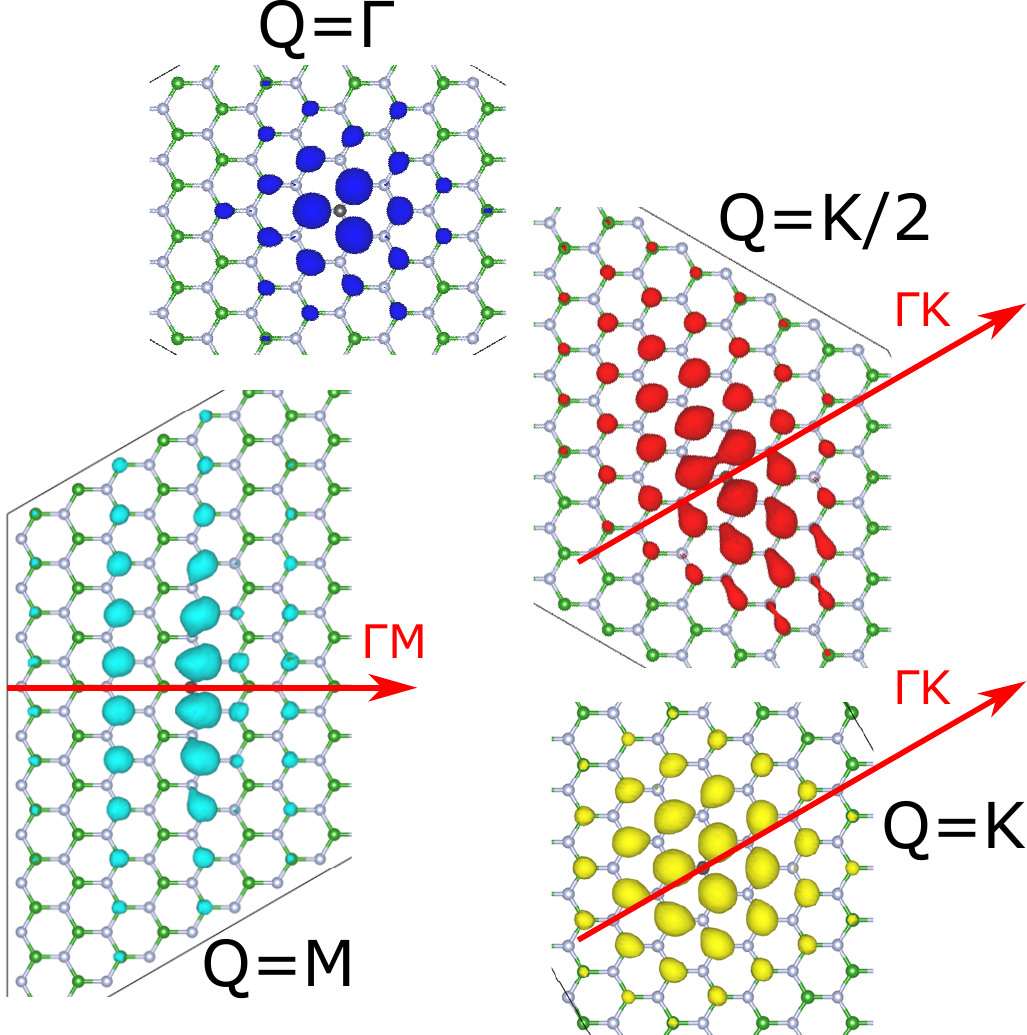}
\caption{\emph{Ab initio} electronic density ($|\Phi_\QQ(\bm{r}_e,\bm{r}_h)|^2$) with fixed hole on a N atom at different exchanged momenta in the AA' bulk phase. At finite $\QQ$, a dashed black line indicates a direction parallel to $\QQ$.}
\label{fig:bulk_exc_wf}
\end{figure}

In order to visualize the distortion of the lowest-energy exciton along its dispersion curve, we plotted the \emph{ab initio} electronic part of the exciton density $|\Phi_\QQ(\bm{r}_e,\bm{r}_h)|^2$ as a function of $\bm{r}_e$ having fixed $\bm{r}_h$ on an arbitrary N atom. The electronic densities at $\QQ=\Gamma$, $\QQ=K/2$, $\QQ=K$ and also $\QQ=M$ are reported in Figure~\ref{fig:bulk_exc_wf}.
The data have been obtained by solving the GW-BSE calculation with the \textsc{yambo} code\cite{yambo} on appropriate supercells such that the desired $\QQ$ point is folded onto $\Gamma$. More details on this method to investigate dispersions at finite $\QQ$ will be available in a future work.
It is worth stressing the similarity of these plots with the corresponding TB plots of the first excitons in the monolayer (Figure~\ref{fig:ml1_exc_disp}). This is a consequence of the fact that even in the bulk the first exciton is basically in-plane, so very similar to the first exciton of the monolayer.
%\footnote{the excitonic wave functions shown in Fig. \ref{fig:10} were computed using the \textsc{yambo} code\cite{yambo}}

\begin{figure*}
\centering
\includegraphics[width=\textwidth]{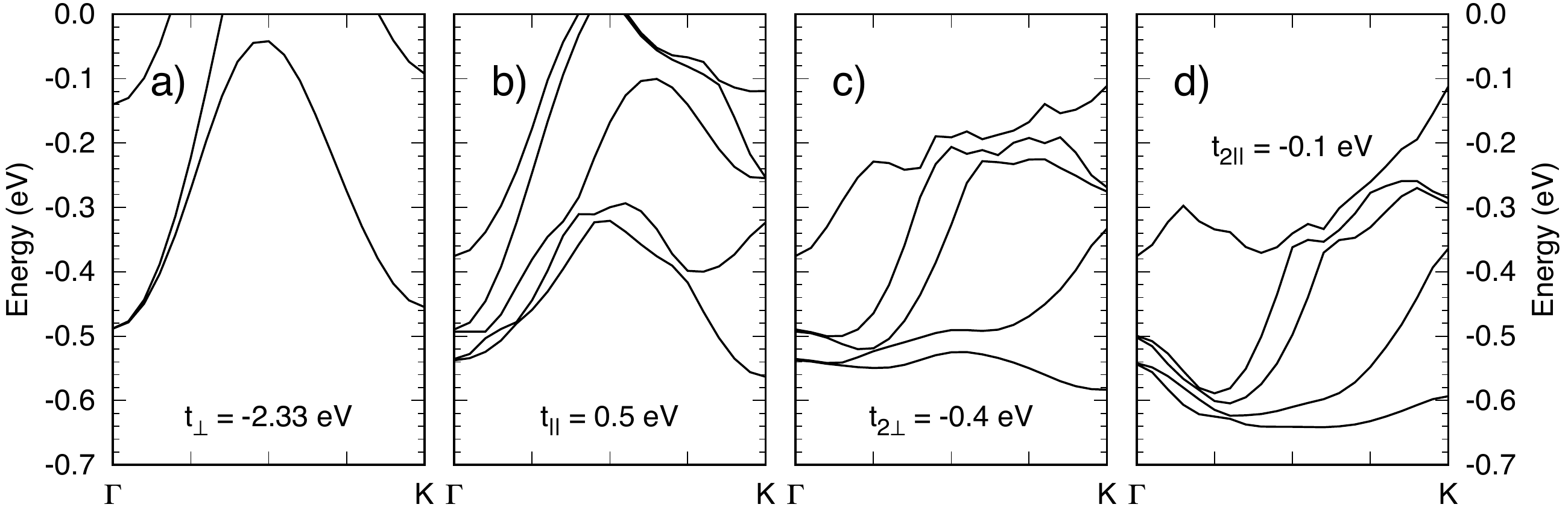}
\caption{Tight-binding dispersion along $\Gamma K$ in the AA' stacking. The $1nn$ and $2nn$ hopping terms $t_{\perp}$, $t_{\parallel}$, $t_{2\perp}$, and $t_{2\parallel}$ are switched on sequentially from panel (a) to panel (d).}
\label{fig:tb_aap_excitons}
\end{figure*}

In the other panels of Figure \ref{fig:excitons_stacking}, we report the corresponding free-carrier and excitonic dispersions (GW-IP and GW-BSE respectively) in the ABC and the AB stacking. In either system, one recognizes the shape of the bottom conduction band in the GW-IP dispersion. In the ABC stacking the double dip is not observed, consistently with the absence of a local minimum of the conduction band at $K$. We remember also that the lowest direct gap of 5.75~eV is not in the high-symmetry lines of the hexagonal cell (so it does not coincide with the entry of Table~\ref{tab:gap_stacking}). The dispersion in the ABC structure passes from 0.38~eV, in the GW-IP case, to 0.14~eV in the GW-BSE case, reproducing the same flattening observed in the AA' case. In this respect, the optical properties of the ABC stacking are predicted to be quite similar to those observed in the AA' case except for the smaller band gap.

More interesting is the prediction of the exciton dispersion in the AB phase. In this material the dispersion of the conduction band is predicted to be very weak (see Table~\ref{tab:bands_stacking} and Figure \ref{fig:bulk_bands-full}), so the electron-hole band flattening is expected to have a stronger impact on the exciton dispersion.  This is indeed the case: the exciton dispersion is reduced from 0.12~eV in the free-carrier picture to -0.02~eV in the exciton picture, which implies an inversion of the nature of the dispersion.  While the indirect nature of the band gap is reflected in the free-carrier dispersion, at the GW-BSE level the lowest exciton is at $\mathbf{Q}=0$, corresponding to a direct exciton, observable in optics.  This implies that in the AB stacking the peaks of luminescence and absorption spectra are expected to coincide, at variance with the other two stackings, and one can expect a stronger luminescence in AB due to the direct transition not mediated by phonons.

Before going deeper, we refer to two appendices where the robustness of these results has been checked. In Appendix~\ref{app:c670} these results have been checked against variations of the inter-plane distance $c$, while in Appendix~\ref{app:gpaw-abinit} different approximations have been used at the DFT and the quasiparticle level.

\subsubsection{Tight-binding model}

We use now the TB model to analyze in detail the influence of the different parameters governing the interplane couplings. We begin with the AA' stacking. To take into account the effective anisotropy of the interactions, we use Coulomb potentials $1/(\epsilon_j R)$ with effective dielectric constants $\epsilon_\perp=6$ within the planes and $\epsilon_\parallel=4.5$ between the planes and cutoff radii of 5 \AA\  and 4.5 \AA,  respectively. These values have been adjusted so as to reproduce reasonably well the first \textit{ab initio} excitons. In particular the order of magnitude of the binding energy of the lowest exciton is now about 0.5-0.6 eV instead of nearly 2 eV for the single layer (see Figure~\ref{fig:tb_aap_excitons}a).

We discuss the role of the different hopping terms by introducing them gradually from the monolayer picture. 
In Figure \ref{fig:tb_aap_excitons}(a) we keep only the $1nn$ hopping $t_\perp$ term, equal to -2.33 eV. The corresponding exciton dispersion is that of the monolayer, but each curve is doubly degenerate because there are two single layers per unit cell. This also means that (without including more distant neighbor hoppings) at $\QQ=0$ the first exciton would be four-fold degenerate.
We should stress here that the dispersion reported in Figure \ref{fig:tb_aap_excitons}(a) differs from that of Figure~\ref{fig:ml1_exc_disp} because of the different potential used in the two calculations: screened 3D in the former, Keldysh in the latter.
In Figure \ref{fig:tb_aap_excitons}(b) the interaction between planes is switched on, i.e. $t_\parallel=0.5$ eV, and the layers start being coupled. As a result the degeneracy of either of the two excitons is lifted along the entire $\Gamma K$ line and at $\QQ=0$ the exciton splits into a couple of doubly degenerate excitons (Davydov splitting).
The electron-hole symmetry must be broken in order to have an indirect electronic gap, so we switch on $t_{2\perp}=-0.4$ in Figure \ref{fig:tb_aap_excitons}(c). As expected, the influence of this parameter is very strong; in particular the lowest exciton branch becomes nearly flat. In Figure \ref{fig:tb_aap_excitons}(d) a further improvement is obtained by accounting for inter-plane $2nn$ integrals $t_{2\parallel}=-0.1$ eV which, in agreement with the \textit{ab initio} results, make appear the indirect minimum of the lowest exciton by breaking the electron-hole symmetry of the TB model.

\begin{figure}[b]
\centering
\includegraphics[width=\linewidth]{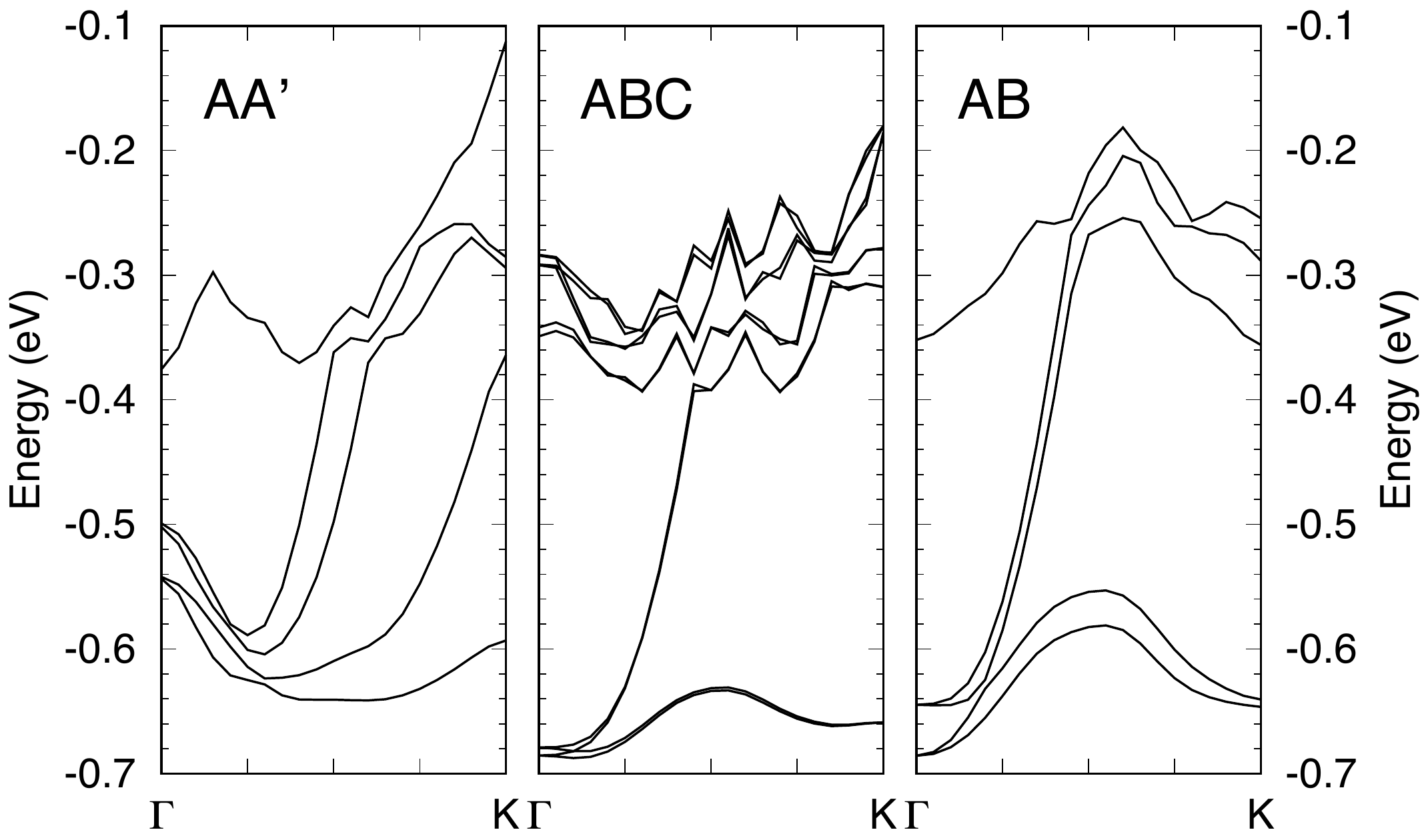}
\caption{Excitonic dispersion along $\Gamma K$ for the three different stackings, AA', ABC and AB, from left to right.}
\label{fig:stack}
\end{figure}

Let us now compare the different stackings, using the same parameters as for the AA' and including all hopping terms. The resulting exciton dispersions are reported in Figure \ref{fig:stack}. It can be seen that both AB and ABC stackings show a ``direct'' lowest exciton. This is not so surprising since the inter-plane coupling geometries are fairly similar in these two cases and quite different from that of AA' (cfr. Figure~\ref{fig:structure}). We recover the conclusion that both the inter-plane coupling and the electron-hole asymmetry play crucial roles to account for the direct-indirect excitonic transition. 

The general agreement with \textit{ab initio} data is still pretty good. Indeed, differences are in the range of a few tens of meV, which is remarkable considering the complexity of the BSE calculations, but still not sufficient for reliable description of optical measurements. 
It is worth pointing out that the quality of the agreement is much poorer with respect to the single-layer case.
We believe that the reason for this lies in the different treatments of electronic screening, and in particular in its momentum-dependence. In \textit{ab initio} calculations this is included through a $\QQ$-dependent RPA dielectric function and similarly in the TB calculation of the monolayer the Keldysh potential is indeed $\QQ$-dependent. Instead, in the TB model of the bulk we used a much rougher approximation, distinguishing parallel from perpendicular screening by means of $\QQ$-independent constants.
Further studies to improve the bulk TB model are currently undertaken in this direction.
%In the monolayer case, it has already been noticed that the use of the Keldysh potential greatly improves the accuracy of the TB-Wannier model because it duly accounts for the $\QQ$-dependence of screening in 2D. The same should be true in bulk systems. Further studies in this direction are currently undertaken.
%Another problem already discussed previously is that the \textit{ab initio} results concerning ABC show surprising differences with other stackings, concerning the values of the GW gap and of the exciton bounding energy. 

\section{Conclusions}
We provide a thorough study of the properties of single-particle and two-particle excitations in hBN monolayer and in three bulk polymorphs: the AA', the ABC, and the AB stackings. We report the first \textit{ab initio} calculations of the exciton dispersion in the AB and the ABC stacking. 
Moreover we devise a tight-binding model for the characterization and the analysis of the excitonic dispersion and wave functions in this material. Using these two theoretical approaches we highlight the impact of interlayer interactions on excitonic properties of hBN.

In the monolayer, our \emph{ab initio} calculations of the band structure and the exciton dispersion along $\Gamma M$ are in agreement with previously published data.\cite{Cudazzo2016,Galvani2016,Koskelo2017} We also provide the exciton dispersion along $\Gamma K$, relevant to discuss the optical properties of the single-layer. We found a very flat dispersion of the lowest energy exciton and the expected linear dispersion of the second exciton at small $\QQ$. Moreover we have derived a tight-binding model for the propagating electron-hole pair which includes kinetic, Coulomb and exchange terms in a Bethe-Salpeter-like formalism. With this model we have highlighted the importance of second-nearest-neighbour hopping terms to describe the dispersion of valence and conduction states and we have been able to analyse the symmetry and image the wave function of the electron-hole pair.

Concerning the bulk, beside summarising some recent results on the AA' stacking,\cite{Schue2018,Sponza2018}  we produce some predictions about the ABC and the AB stacking. In the ABC phase, the exciton properties are predicted to be similar to the more common AA' phase, in particular in relation with the exciton dispersion.  An intriguing result is the prediction of a band gap around 0.5~eV smaller than the AA' one. There are indications that this has to be ascribed to specific screening properties of the ABC stacking distinct from those of the other two bulk phases, but we foster for more investigations on this subject.

We generalised our tight-binding model to the three bulk systems and we pointed out the importance of the second-nearest-neighbour in-plane and inter-plane hopping terms. As their introduction breaks the electron-hole symmetry, they are essential to reproduce the indirect gap and the dispersion of the lowest energy excitons. Also, we indicate a route to improve this model by including appropriate $\QQ$-dependent effective dielectric functions.

Our main finding is the peculiar exciton dispersion in the AB stacking. In this material our \textit{ab initio} simulations point out a striking difference between the single-particle and the two-particle excitations. At the single-particle level (band structure), the material exhibits an indirect gap of $\sim$6.1~eV (probably underestimated of about 0.5~eV~\cite{Sponza2018}), but at the two-particle level (exciton dispersion) the material is predicted to have a direct gap.  This is due to  the strong momentum dependence of the exciton binding energy $E_b(\mathbf{Q})$, which is approximately halved when passing from $\mathbf{Q}=\Gamma$ to $\mathbf{Q}=1$~\AA$^{-1}$. This reduction is observed also in the other two materials, but in the AB stacking the variation of $E_b(\mathbf{Q})$ is larger than the dispersion of the band structure resulting in a direct exciton transition despite the indirect band gap. This finding will have strong implication for luminescence, that we expect to be much stronger in the AB than in the other bulk phases.

\acknowledgments
The authors want to acknowledge Dr. J. Barjon of UVSQ-CNRS, Universit\'{e} Paris-Saclay.
The French National Agency for Research (ANR) is acknowledged for funding this work under the project GoBN (Graphene on Boron Nitride Technology), Grant No. ANR-14-CE08-0018.
F. Paleari, T. Galvani and L. Wirtz acknowledge support from the National Research Fund, Luxembourg (Projects EXCPHON/11280304 and INTER/RCN/16/11402387/2Defect).
The research leading to these results has received funding from the European Union Seventh Framework Program under grant agreements no. 696656 Graphene Core1 and no. 785219 Graphene Core2.

%%%%%%%%%%%%%%%%%%%%%%%%%%%%%%%%%

\appendix

%%%%%%%%%%%%%%%%%%%%%%%%%%%%%%%%%%%%%%%%%%%%

\section{Tight-binding band structure for single layer and bulk hBN systems\label{app:formulae}}

In this section we report the exact and approximated formula for the systems studied, together with a summary of the parameters used. 

In the monolayer the exact diagonalization of TB Hamiltonian leads to expression
\begin{equation}
E_{\kk_\perp\, e(h)} = \pm \sqrt{ \Delta^2 + t_\perp^2|\gamma(\kk_\perp)|^2 } + t_{2\perp}\left(|\gamma(\kk_\perp)|^2-3\right)
\label{eq:ml_exact_formula}
\end{equation}
where the signs ($+$) and ($-$) are for conduction and valence states respectively. 
The approximate expression~\eqref{eq:TB_single-layer_1particle} valid for $\Delta \gg t_{\perp}^2|\gamma(\kk_\perp)|^2$ holds very well in the $KM$ line.
The optimization of the parameters $\Delta=3.625$~eV, $t_\perp=-2.33$~eV, and $t_{2\perp}=-0.4$~eV have been done manually by comparing these exact expressions with the \emph{ab initio} bands in the $KM$ region.
All the analysis on the single-particle band structure and Figure~\ref{fig:ml1_bands}b come from the exact expressions above. Instead, in deriving the excitonic model, it has been necessary to adopt the approximate expressions reported in the main text in order to truncate the Wannier functions to the first-nearest-neighbours ($1nn$). Given the range of validity of the approximate formula and the agreement shown with the \emph{ab initio} results, this approximation is widely justified.

In the AA' stacking, the exact expressions including second-nearest-neighbour ($2nn$) hopping terms are given below for electron states:
\begin{equation}
\begin{split}
E_{\kk\,e}^{AA'} &=  \sqrt{\Delta^2 + \left[ t_\perp|\gamma(\kk_\perp)| \pm 2t_\parallel\cos(k_\parallel c) \right]^2 } + \\
& \quad + t_{2\perp}(|\gamma(\kk_\perp)|^2 - 3) \pm 2 t_{2\parallel}|\gamma(\kk_\perp)|\cos(k_\parallel c)\, , \label{eq:aa_formula_electrons}
\end{split} 
\end{equation}
and hole states:
\begin{equation}
\begin{split}
E_{\kk\,h}^{AA'} &=  - \sqrt{\Delta^2 + \left[ t_\perp|\gamma(\kk_\perp)| \pm 2t_\parallel\cos(k_\parallel c) \right]^2 } + \\
& \quad + t_{2\perp}(|\gamma(\kk_\perp)|^2 - 3) \pm 2 t_{2\parallel}|\gamma(\kk_\perp)|\cos(k_\parallel c)\,. \label{eq:aa_formula_holes}
\end{split}
\end{equation}
The approximate equation~\eqref{AA'bands} reported in the main text includes only $1nn$ terms.
While $\Delta$, $t_\perp$ and $t_{2\perp}$ have been kept as in the monolayer, the values $t_{\parallel}=0.5$~eV and $t_{2\parallel}=-0.1$~eV have been fitted comparing these exact expressions with \emph{ab initio} results.
In the same way, Figure~\ref{TBbands} has been obtained from these exact expressions.
Table~\ref{tab:tb_parameters} collects the values of the optimized parameters used in our calculations.

\begin{table}
\centering
\bgroup
\setlength\tabcolsep{3mm}
\begin{tabular}{ccccc}
\hline \hline
$\Delta$ & $t_\perp$ & $t_{2\perp}$ & $t_\parallel$ & $t_{2\parallel}$ \\
\hline
3.625 & -2.33 & -0.4 & 0.5 & -0.1 \\
\hline \hline
\end{tabular}
\egroup
\caption{Optimal TB parameters. Energies in eV}
\label{tab:tb_parameters}
\end{table}

%In some cases the sign of the hopping terms is irrelevant, for instance for $t_\perp$. However in many cases a ``natural" sign of the hopping integral can be derived from geometrical arguments that are sketched in Figure~\ref{fig:hopping_sign}. Consider the symbolic notation  $t \propto \langle p_z | U | p_z \rangle$ with $U$ appropriate potential. Given $U<0$ and taking into account the amplitude of the $p_z$ orbitals, one gets that $t_\perp<0$, $t_{2\perp}<0$, and $t_\parallel>0$.
%The same arguments would suggest $t_{2\parallel}>0$, but positive values actually spoil the agreement with the \emph{ab initio} results. This is not so surprising though, since the $t_{2\parallel}$ hopping term is definitely the 

\section{Tight-binding exciton model at $\QQ\neq 0$ (with no exchange contributions)}
\label{TBdisp}

A general exciton state $\ket{\Phi_{\bm{Q} }}$ is obtained by combining electron and hole one-electron states of wave vector $\bm{k}_e=\bm{k}$ and $\bm{k}_h = \bm{k}-\bm{Q}$, respectively:
\begin{equation}\begin{split}
\ket{\Phi_{\bm{Q} }} &= \sum_{\bm{k}}\Phi_{\bm{kQ}} \ket{\bm{k}eh,\bm{Q}}\\
\ket{\bm{k}_e,\bm{k}_h} \equiv \ket{\bm{k}eh,\bm{Q}} &= a^\dagger_{\bm{k}e} a^{}_{\bm{k}-\bm{Q}h} \ket{\emptyset}  \;  ,
\end{split}\end{equation}
where $\Phi_{\bm{kQ}}$ is the exciton wave function in the $(\bm{k},\bm{Q})$ representation. We can also define elementary excitonic states in real space:
\begin{equation}\begin{split}
\label{R}
\ket{\bm{R}eh,\bm{Q}} &=  \frac{1}{\sqrt{N}}\,
\sum_{\substack{\bm{n},\bm{m} \nonumber \\
\bm{m}-\bm{n}=\bm{R}}} e^{-i\bm{Q}.\bm{n}} a^\dagger_{\bm{m},e} a_{\bm{n},h^{}}        \ket{\emptyset} \\
&=\frac{1}{\sqrt{N}}\, \sum_{\bm{n}}e^{-i\bm{Q.n}} 
a^\dagger_{\bm{n}+\bm{R},e} \; a_{\bm{n} ,h}^{}  \ket{\emptyset}  \;,
\end{split}
\end{equation}
with $N$ the number of sites. In the expression above, $\bm{m}\in \Lambda_B$ are boron sites and $\bm{n}\in\Lambda_N$ are nitrogen sites. If the origin is located on an $\bm{n}$ site, the set of relative distance vectors $\RR=\bm{m}-\bm{n}$ coincides therefore with the $\Lambda_B=\Lambda_N+\bm{\tau}_3$ lattice.\cite{Note1}
The state $\ket{\bm{R}eh,\QQ}$ is the Bloch state describing the motion of an electron-hole pair of size $\bm{R}$.
It can be checked that:
\begin{equation}
\ket{\bm{k}eh,\QQ} = \frac{1}{\sqrt{N}} \sum_{\bm{R}}\, e^{i\bm{k}\cdot{\RR}}\ket{\bm{R}eh,\QQ} \; .
\end{equation}

For sake of simpliciy, in the following we will frequently drop the $eh$ label.
The kinetic part of the excitonic hamiltonian is obtained from the difference of the single particle hamiltonians, $H^0_{eh} = H^0_{e} - H^0_{h}$. For the electron part the action of the hamiltonian is given by:
\begin{equation}
H^0_{e} \ket{\bm{R},\QQ} = \sum_{\rho_e} h_e(\bm{\rho}_e) \, \ket{\bm{R}+\bm{\rho}_e,\QQ} \; ,
\end{equation}
where $h_e(\bm{\rho}_e)$ is the hopping integral associated with the vector $\bm{\rho}_e$ connecting two conduction (boron) sites. In the same way the action of the hole hamiltonian is given by:
\begin{equation}
H^0_{h} \ket{\bm{R},\QQ} = \sum_{\bm{\rho}_h} h_h(\bm{\rho}_h)e^{-i\bm{Q}\cdot\bm{\rho}_h} \, \ket{\bm{R}-\bm{\rho}_h,\QQ}.
\end{equation}

In the monolayer, the vector sets $\{\rho_e\}$ and $\{\rho_h\}$ are identical, and for the simple model used at $\QQ=0$,\cite{Galvani2016} $h_h(\bm{\rho}_h) = -h_e(\bm{\rho}_e)=t^2_\perp/2\Delta = t_{ex}/2$, so that finally, 
\begin{equation}
H^0_{eh} \ket{\bm{R},\QQ} = \sum_{\rho_e}\left( h_e(\bm{\rho}) \, - h_h(\bm{\rho})e^{i\bm{Q} \cdot \bm{\rho}} \right)\, \ket{\bm{R}+\bm{\rho},\QQ} \; ,
\label{hopQ}
\end{equation}
and the effective hopping integral between $1nn$ at finite $\QQ$ becomes:
\begin{equation}
\begin{split}
\bra{\bm{R}',\QQ} H^0_{eh} \ket{\bm{R},\QQ} &= \left( h_e(\bm{R}'-\bm{R}) \, - h_h(\bm{R}'-\bm{R})  e^{i\bm{Q}\cdot (\bm{R}'-\bm{R})} \right) \nonumber\\
&= \frac{t_{ex}}{2} \left(1+ e^{i\bm{Q}\cdot(\bm{R}'-\bm{R})} \right) \; ,
\end{split}
\end{equation}
which reduces to $t_{ex}$ when $\bm{Q}=0$, as expected. The diagonal part of the hamiltonian does not depend on $\bm{Q}$ and is therefore the same as before. Coming back to Eq.(\ref{hopQ}) it is not difficult to include $2nn$ and even to introduce different hopping integrals between boron sites and between nitrogen sites.

Finally, let $\Phi_{\RR\QQ}=\braket{\RR,\QQ}{ \Phi }$ be the exciton wave function in the $(\bm{R},\bm{Q})$ representation, then the Bethe-Salpeter-Wannier (BSW) equation without the exchange term,  becomes:
\begin{equation}
E\, \Phi_{\bm{RQ}} = \sum_{\bm{R}'} \bra{\bm{R},\QQ} H^0_{eh} \ket{\bm{R}',\QQ}\Phi_{\bm{R'Q} }+  \Xi_{\bm{R}} \Phi_{\bm{RQ}} 
\end{equation}
where we have used the fact that $\bra{\RR,\QQ}\Xi\ket{\RR'\QQ}$ is diagonal in $\RR$ (cfr. Equation~\eqref{eq:Coulomb}).

Actually, the choice of the definition of $\bm{k}$ from the pair $(\bm{k}_h,\bm{k}_e)$ is not unique. Instead of the pair $(\bm{k-Q},\bm{k})$, we could have chosen the pair $(\bm{k-Q/2},\bm{k+Q/2})$. This should of course not modify the eigenvalues of the BSW equation, but modifies the phase of the ``real space'' state $\ket{\bm{R}eh,\bm{Q}}$ defined in Eq. (\ref{R}). In particular, in the latter gauge, the effective hopping integral $\bra{\bm{R},\QQ} H^0_{eh} \ket{\bm{R}',\QQ}$ becomes real, which may be convenient in some cases. Finally, let us calculate the full wave function in real space:
\begin{widetext}
\begin{equation}
\begin{split}
\Phi_\QQ(\bm{r}_h,\bm{r}_e) &=  \sum_{\bm{R}}\Phi_{\bm{RQ}} \, \braket{\bm{r}_h,\bm{r}_e}{\bm{R},\bm{Q}} \\
&=\frac{1}{\sqrt{N}}\,\sum_{\bm{R}}\Phi_{\bm{RQ}}\; \sum_{\bm{n}}e^{-i\bm{Q}\cdot\bm{n}} \phi_e(\bm{r}_e-\bm{n}-\bm{R}) \phi_h(\bm{r}_h-\bm{n})  \; .
\end{split}\end{equation}
\end{widetext}

Remembering that we fixed the hole position on the valence site at the origin, $\bm{r}_h =\bm{n}= 0$, so that:
\begin{equation}
\Phi_\QQ(\bm{r}_h=0,\bm{r}_e) \propto \sum_{\bm{R}}\Phi_{\bm{RQ}}\; \phi_e(\bm{r}_e-\bm{R}) .
\end{equation}
Actually $\bm{r}_h$ should be taken slightly above $\bm{n}$ since the $\pi$ wave functions $\phi(\bm{r})$ vanish at the origin. Since the $\phi(\bm{r}_e -\bm{R})$ are localized on sites $\bm{R}$ (and also a little bit on their neighbours if genuine Wannier functions are used), we see that the full wave function can be (partially) represented as the superposition of localized contributions weighted by the amplitudes $\Phi_{\bm{RQ}}$.  Notice also that the chosen gauge ensures that the wave function is invariant when $\bm{Q}$ is replaced by $\bm{Q}+\bm{G}$, where $\bm{G}$ is a reciprocal lattice vector. The previous formalism can be extended to bulk stackings by introducing interlayer hopping integrals and extending Eq.(\ref{hopQ}) accordingly.

\section{Exchange contributions}
\label{exchange}

Using the tight-binding Bloch states, the exchange kernel $\bra{\bm{k}'_e,\bm{k}'_h}V\ket{\bm{k}_e,{\bm{k}_h}}$ involved in the Bethe-Salpeter equation can be written:
\begin{widetext}
\begin{equation}
\begin{split}
\bra{\bm{k}'_e,\bm{k}'_h}V\ket{\bm{k}_e,{\bm{k}_h}} &= \int d\bm{r} d\bm{r'} \phi^*_{k'e}(\bm{r}')  \phi_{k'h}(\bm{r}') \frac{2e^2}{|\bm{r}-\bm{r}'|}\,
\phi_{ke}(\bm{r})  \phi^*_{kh}(\bm{r}) \\
&= \frac{1}{N^2} \sum_{\bm{n},\bm{m},\bm{n}',\bm{m}'} 
\exp\{  i[\bm{k}\cdot\bm{n} -  (\bm{k} - \bm{Q})\cdot\bm{m} - \bm{k}'\bm{n}' + (\bm{k} - \bm{Q}')\cdot\bm{m}' ]\} \times  \\
&\quad \quad\times  \int d\bm{r} d\bm{r'} \varphi_{e}(\bm{r}'-\bm{n}')  \varphi^*_{h}(\bm{r}'-\bm{m}') \frac{2e^2}{|\bm{r}-\bm{r}'|}\,
\varphi_{e}(\bm{r}-\bm{n})  \varphi^*_{h}(\bm{r}-\bm{m})   \; .
\end{split}
\end{equation}
\end{widetext}

Translation invariance implies $\bm{Q}=\bm{Q'}$. Then the most important integrals are those where all involved lattice sites are as closest to each other as possible. In usual treatments it is assumed that they are all identical, but here we know that the $\bm{n}$ and $\bm{m}$ sites belong to distinct triangular sublattices, so that the best we can do is to assume that $\bm{n}-\bm{m} = \bm{\tau}$ and $\bm{n}'-\bm{m}' = \bm{\tau}'$, where $\bm{\tau}$ and $\bm{\tau}'$ are the shortest vectors joining the conduction sites to the fixed hole site. At least this is true to lowest order in $t_\perp/2\Delta$, in which case the Wannier functions are localized atomic orbitals centred on the lattice sites (see Ref. [\onlinecite{Galvani2016}] for a discussion). Coming back to a real space discussion it is then found that $V$ couple states $\ket{\bm{\tau},\QQ}$ and $\ket{\bm{\tau}',\QQ}$:
\begin{equation}
\bra{\bm{\tau}',\QQ} V \ket{\bm{\tau},\QQ} =  \sum_{\bm{\rho}} J_{\bm{\tau}'\bm{\tau}}(\bm{\rho)}
e^{-i\,\bm{Q} \cdot \bm{\rho}} \; ,
\label{eq:exchange-term}
\end{equation}
with :
\begin{equation}
J_{\bm{\tau}'\bm{\tau}}(\bm{\rho)}=\int d\bm{r} d\bm{r}\,'
\frac{ \varphi_e(\bm{r}-\bm{\tau}) \varphi_h(\bm{r}) 2e^2
\varphi_e(\bm{r}\,'-\bm{\tau}\,')
\varphi_h(\bm{r}\,') }{|\bm{r}\,'-\bm{r} + \bm{\rho}|}   \; .
\end{equation}

%
%\begin{equation}
%\begin{split}
%&J_{\bm{\tau}'\bm{\tau}}(\bm{\rho)}=\\
% &=\int d\bm{r} d\bm{r}\,'
%\frac{ \varphi_e(\bm{r}-\bm{\tau}) \varphi_h(\bm{r}) 2e^2
%\varphi_e(\bm{r}\,'-\bm{\tau}\,')
%\varphi_h(\bm{r}\,') }{|\bm{r}\,'-\bm{r} + \bm{\rho}|}   \; .
%\end{split}\end{equation}
%

Then we use the Fourier development of  $1/r =\sum_{\bm{q}} (4\pi /\Omega q^2) e^{i\bm{q}\cdot{\bm{r}}}$, where $\Omega$ is the volume, and then:
\begin{equation}
J_{\bm{\tau}'\bm{\tau}}(\bm{\rho)} = \sum_{\bm{q}} \frac{8\pi e^2}{\Omega q^2} e^{i\,\bm{q}\cdot\bm{\rho}} M(\bm{\tau},\qq)^* M(\bm{\tau}',\qq)\;,
\label{eq:J-full}
\end{equation}
having introduced the matrix elements $M(\bm{\tau},\qq) = \int d\bm{r} \,\varphi_e(\bm{r}-\bm{\tau}) \varphi_h(\bm{r}) e^{i\bm{q} \cdot \bm{r}}$  and remembering that $\varphi$ are real functions.

Let us now calculate the integrals $M$.
To be consistent they should be neglected in the simplest tight-binding model, but actually the $\varphi(\bm{r})$ are here Wannier functions with components on the neighbouring sites,
\begin{equation}
\begin{split}
&\varphi_e(\bm{r}) \simeq  \varphi^{\text{at}}_e(\bm{r}) - \frac{t_\perp}{2\Delta} \sum_{\bm{\tau}} \varphi^{\text{at}}_h(\bm{r}+\bm{\tau}) \text{ and }\\
&\varphi_h(\bm{r}) \simeq  \varphi^{\text{at}}_h(\bm{r}) + \frac{t_\perp}{2\Delta} \sum_{\bm{\tau}} \varphi^{\text{at}}_e(\bm{r}-\bm{\tau}) 
\end{split}
\end{equation}
where the $\varphi^{\text{at}}_j(\bm{r})$ are the genuine localized atomic orbitals (see Ref. [\onlinecite{Galvani2016}] for a discussion). As a consequence  in the macroscopic limit $\bm{q} \to 0$  we can write:
\begin{equation}
M(\bm{\tau},\qq)     \simeq \frac{t_\perp}{2\Delta} (e^{i\bm{q} \cdot \bm{\tau}} - 1)  \; ,
\label{eq:M-terms}
\end{equation}
which clearly corresponds to dipolar integrals, i.e. overlap integrals weighted by $e^{\pm i\bm{q} \cdot \bm{r}}$.

Finally, using \eqref{eq:M-terms}  and \eqref{eq:J-full} into \eqref{eq:exchange-term} one gets
\begin{equation}
\begin{split}
\bra{\bm{\tau}',\QQ} V \ket{\bm{\tau},\bm{Q}} & =  \sum_{\bm{G}} \frac{8\pi e^2}{\Omega_{\text{at}} |\bm{Q}+\bm{G}|^2} \left(\frac{t_\perp}{2\Delta}\right)^2 \times  \\
&  \quad \times   \left(e^{-i(\bm{Q}+\bm{G}) \cdot \bm{\tau}} - 1 \right) \left(e^{i(\bm{Q}+\bm{G}) \cdot \bm{\tau}'} - 1\right) \;,
\end{split}\end{equation}
where $\Omega_{\text{at}}$ is the volume of the unit cell,  and $\bm{G}$ are reciprocal lattice vectors of a system with periodic boundary conditions. This derivation is actually similar to the one  by Qiu et al.\cite{Qiu2015} carried out within a $\bm{k}\cdot\bm{p}$ formalism and also to the result obtained by Cudazzo et al.\cite{Cudazzo2016} with a simplified model for the electronic structure not based on TB formalism. As derived, the sum over $\bm{G}$ is done on three dimensions (3D), but in the two-dimensional (2D) limit, the sum over the $z$ components of $\bm{G}$ transforms into an integral which can be performed analytically. In the remaining sum over in-plane components $\bm{G}_\perp$, only the $\bm{G}_\perp=0$ term is singular in the $\bm{Q} \to 0$ limit, and we finally obtain:
\begin{equation}
\bra{\bm{\tau}',\QQ} V \ket{\bm{\tau},\QQ} \simeq   J   \frac{1}{4Q\tau} 
(e^{-i\bm{Q} \cdot \bm{\tau}} - 1) (e^{i\bm{Q} \cdot \bm{\tau}'} - 1) 
\end{equation}
with $J=4\pi e^2t^2/(3\sqrt{3}\tau\Delta^2)$ where the area of the 2D unit cell is equal to  $3\sqrt{3}\tau^2$ , $\tau = |\bm{\tau}| = a/\sqrt{3}$ being the $1nn$ distance and $a$ the lattice parameter of the triangular lattice.
%\begin{equation}
%J= \left(\frac{t}{\Delta}\right)^2 \frac{4\pi e^2}{3\sqrt{3}\tau}   \; ,
%\end{equation}
%where we have taken into account that 

%%%%%%%%%%%%%%%%%%%%%%%%%%%%%%%%%%%%%%%%%%%%

\section{Perturbative treatment at small $\QQ$ of exchange terms in the monolayer\label{app:perturb}}

Let us evaluate the effect of the exchange $V$ on the first two (degenerate) excitons when one goes slightly away from $\QQ=0$.
At $\QQ=0$ the two degenerate states are labelled $\Phi^+$ and $\Phi^-$ as in the main text.
Let $\theta$ be the polar angle of $\bm{Q}$, then the matrix of $V$ in this $(\Phi^+,\Phi^-)$ space reads
\begin{equation}
V \propto 
\QQ\cdot\bm{d}_\Phi\begin{pmatrix}
1 &e^{2i\theta} \\
e^{-2i\theta} &1\\
\end{pmatrix} \; ,
\end{equation} 
where $\bm{d}_\Phi=\sum_{\bm{\tau}} \bm{\tau} \braket{\bm{\tau}}{\Phi}$ is the dipolar matrix element of the $\ket{\bm{\Phi}}$ state.
The eigenvalues of the matrix above are equal to 0 and 2.
The lower branch (eigenvalue 0) is a ``transverse" eigenstate for which $\bm{Q} \cdot \bm{d}_{\Phi}=0$ at small $\bm{Q}$, so it can be optically active. Moreover its dispersion is not affected by the exchange term.
Instead  the dispersion of the upper branch (eigenvalue 2) is affected by the singular component of the exchange, which is linear in $Q$,\cite{Wu2015,Qiu2015} so the degeneracy is indeed lifted by effect of the exchange interaction. Moreover it is a longitudinal eigenstate, hence dark at normal incidence but possibly active for oblique incidence. Notice that an opposite terminology is sometimes used, for example in electron energy loss spectroscopy the active modes are longitudinal so they are the bright ones.

%%%%%%%%%%%%%%%%%%%%%%%%%%%%%%%%%%%%%%%%%%%%

\section{A toy model for localized excitons: excitons on a Kagome lattice and flat bands}
\label{kagome}

\begin{figure}
\centering
\includegraphics[width=5cm]{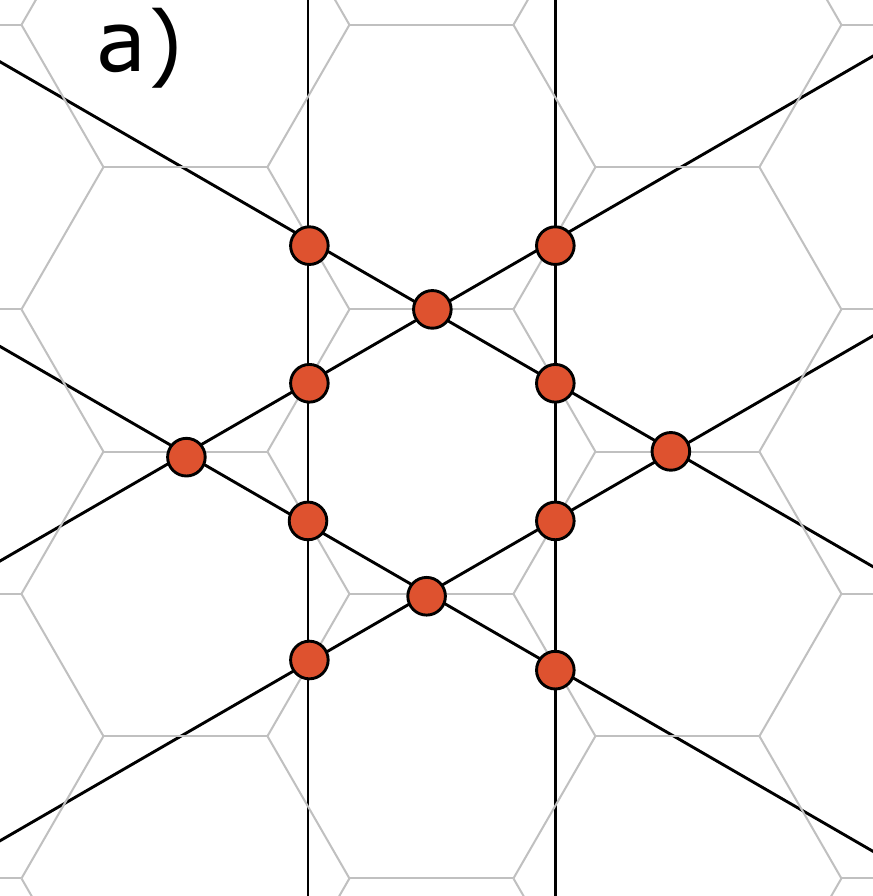} 

\vspace{5mm}

\includegraphics[width=7cm]{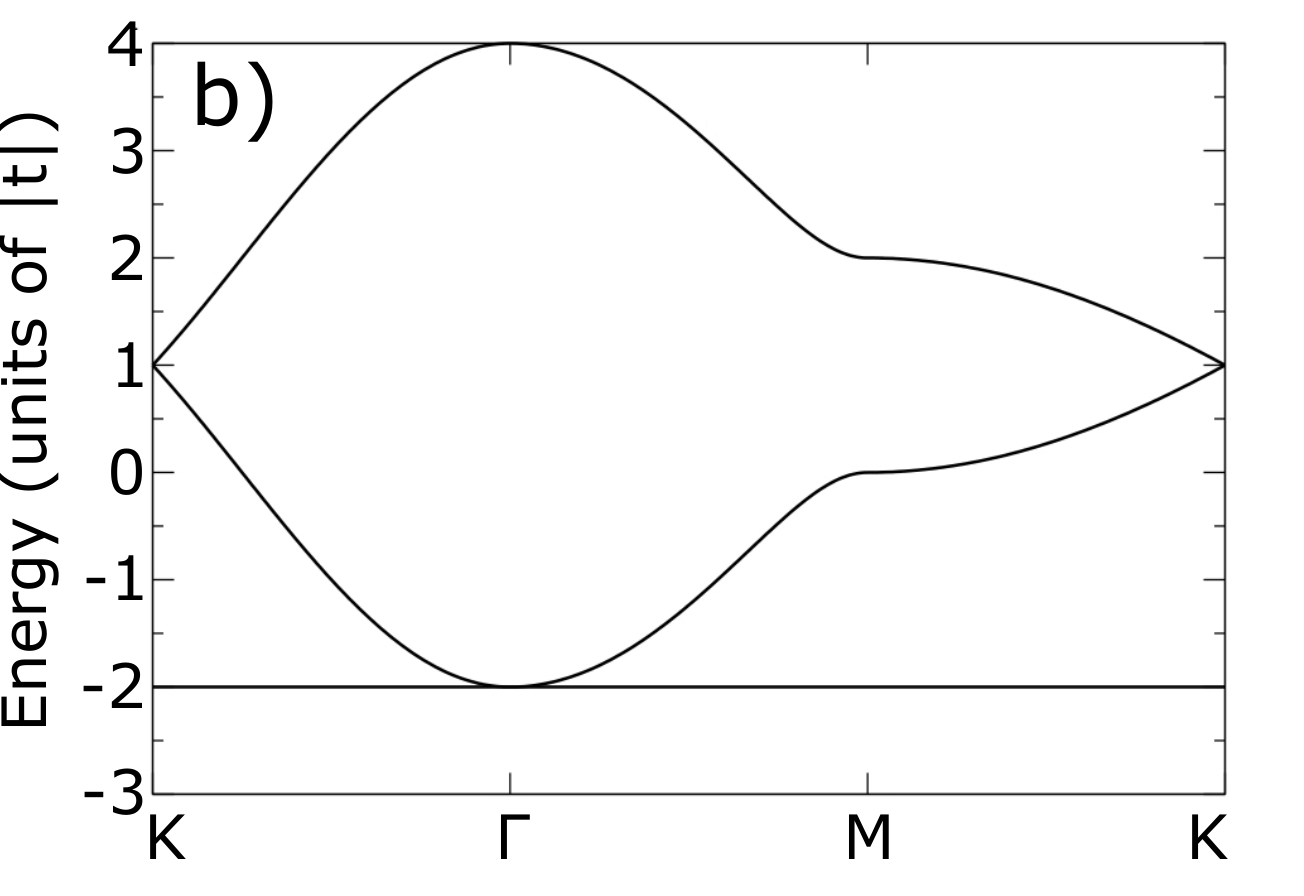} 
\caption{(a) The lattice points at the middle of the bonds of a honeycomb lattice form a Kagome lattice. Each point of this lattice represent an excitonic pair which can jump from one site to a neighbouring one. (b) Band structure of the Kagome lattice. Energy is give in units of the hopping integral $|t|$.}
\label{kagome1}
\end{figure}

We have seen that at low $\bm{Q}$ the degeneracy of the ground state exciton of symmetry $E$ is lifted into transversal and longitudinal states. The upper branch rises linearly with $Q$ due to exchange effects. It is frequently argued that this is the main reason for the splitting of the states. Actually within a simple Wannier-Mott continuous model where the  degenerate states are associated to separate valleys around point $K$ or $K'$ the dispersion of the branches are expected to be similar since the effective masses are identical in both valleys. This is not necessarily the case when intervalley interactions are significant. We show below that the effect can be huge in the limit of strong localization of the exciton. 

Let us assume therefore that the considered exciton wave fonction is confined to B-N $1nn$ pairs. Fixing the position of the hole, we have then three possible orientations of the excitonic pair. In the formalism used up to now, they are labelled by the corresponding vectors $\bm{\tau}$. In the exciton hamiltonian the ``kinetic energy" part related to the free motion of the hole and of the electron, the jump of the exciton from point $\bm{\tau}$ to a neighbouring site is precisely accounted for  within the TB model by the hopping integral $t_{ex}$. But we can also move the hole which will jump from one site to a neighbouring one on its triangular sublattice. Usually it is very difficult to represent both motions of holes and electrons. Here this is possible because of the constraint that they should remain $1nn$ (on the honeycomb lattice). The method is to mark each pair by the position of its centre. All these positions lie in fact on a so-called Kagome lattice where each site has four $1nn$. It is easy to realize then that the motion of an excitonic pair on this lattice corresponds to first neighbour jumps on this Kagome lattice (Fig. \ref{kagome1}(a)). The problem of describing the dispersion of the exciton states has been reduced to a single particle TB band problem on the Kagome lattice with $1nn$ interactions. The on-site matrix element are all the same and equal to  the Coulomb energy of the pair, taken here as the origin of energies.

The solution of this problem is known and shows very interesting features. The dispersion curves are shown in Fig. \ref{kagome1}(b). The Brillouin zone is still the hexagonal one, and there are three branches since  the Kagome lattice  has three sites per unit cell. In particular there is a completely flat band. In general flat bands indicate the presence of localized states. This is of course what is obtained if the interatomic jumps are forbidden. What is surprising here is that jumps are allowed, but a basis of localized states should exist. They do exist and are actually localized on the hexagons of the underlying honeycomb lattice. This has been discussed in many places. Flat bands may produce surprising effects as easy self-localization of extended states in the presence of small perturbations. For a review see for instance Ref. [\onlinecite{Zheng2014}]; see also Ref.[\onlinecite{Barreteau2017}]. It is tempting to apply this model to the behaviour of our ground state excitons by looking at the two lowest states of the Kagome lattice. The third and highest level is unphysical in our context since anyway there are other, more extended, excitons at high energy and the continuum of single particle excitations.

By this model we demonstrate that if excitons are very localized a flat band may appear, which undoubtedly will induce peculiar effects. In the case of hBN, although the ground state exciton is fairly localized, we are clearly not in this extreme limit and the difference in dispersion of the two branches at low $\bm{Q}$ is principally due to exchange contributions. This does not mean that the Kagome model is useless, since for example it provides explicit solutions for the energy and the wave functions.

%%%%%%%%%%%%%%%%%%%%%%%%%%%%%%%%%%%%%%%%%%%%

%\section{Additional data on bulk polymorphs\label{app:full-zone_band-structure}}
%
%
%
%
%We provide additional data on the bulk structures that have not been included in the main text. 
%The GW band structures in the full hexagonal Brillouin zone of the three stackings are reported in Figure~\ref{fig:bulk_bands-full}.
%It is possible to appreciate that in the AA' and the AB systems each valence and conduction band on the $AHL$ plane is doubly degenerate.
%These bands disperse as the average of the corresponding pairs on the $\Gamma K M$ plane, as predicted by the TB equations~\eqref{AA'bands} and \eqref{ABbands} (cfr. main text at section IV.A.).

%%%%%%%%%%%%%%%%%%%%%%%%%%%%%%%%%%%%%%%%%%%%
\section{Varying the inter-plane distance\label{app:c670}}

\begin{figure}
\centering
\includegraphics[width=0.48\textwidth]{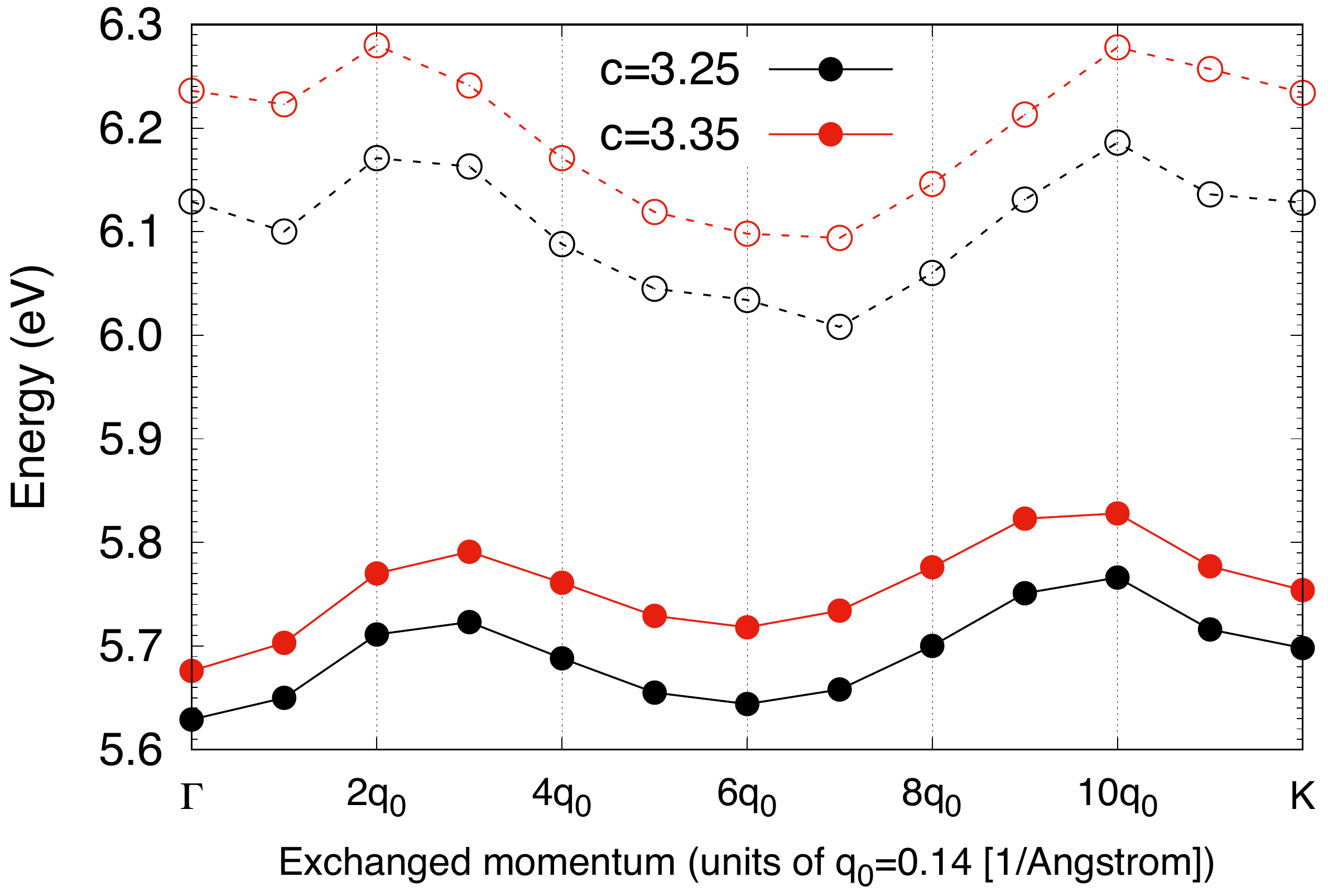}
\caption{Exciton dispersion along $\Gamma K$ (full circles) and GW transitions (open circles) at two different interplane distances $c$ in the AB stacking. The unit $q_0=0.14$\AA$^{-1}$.}
\label{fig:eps_compar}
\end{figure}

We have decided not to perform structural optimization in the bulk structures. For the comparison to be meaningful, we have used the same cell parameters in the three structures. In particular, the interlayer distance $c=3.25\,$\AA\  has been used in the three cases.
However, we have explored in the case of the AB stacking the effect on the band structure and the exciton dispersion of a variation of the interlayer spacing.
In Figure~\ref{fig:eps_compar}, we show that a variation from $c=3.25\,$\AA\  to $c=3.35\,$\AA, which corresponds to an increase of  3\%, does not change the conclusions of the main text.

It can be seen that the changes in the exciton dispersion induced by the variation of $c$ are negligible. 
In particular, the prediction that the nature of the gap changes from indirect to direct when the electron-hole interaction is taken into account is not compromised by this change.
In fact, the value $c=3.35\,$\AA\, is probably closer to the actual interplane distance in this material, so we expect this effect to be larger than what predicted in the main text on the basis of $c=3.25\,$\AA.
We note, by the way, that the difference between IP-transition data of the two sets (distance between dashed curves in the Figure) is larger than the corresponding difference between excitonic data (solid curves).  This can be seen for instance in the $\Gamma$ or in the $K$ points. It indicates the expected trend of an increase of both the IP-transition energies (\textit{i.e.} an increase of the gap) and of the exciton binding energy at higher $c$.

%%%%%%%%%%%%%%%%%%%%%%%%%%%%%%%%%%%%%%%%%%%%

\section{Robustness of the prediction\label{app:gpaw-abinit}}

\begin{figure}[b]
\centering
\includegraphics[width=0.48\textwidth]{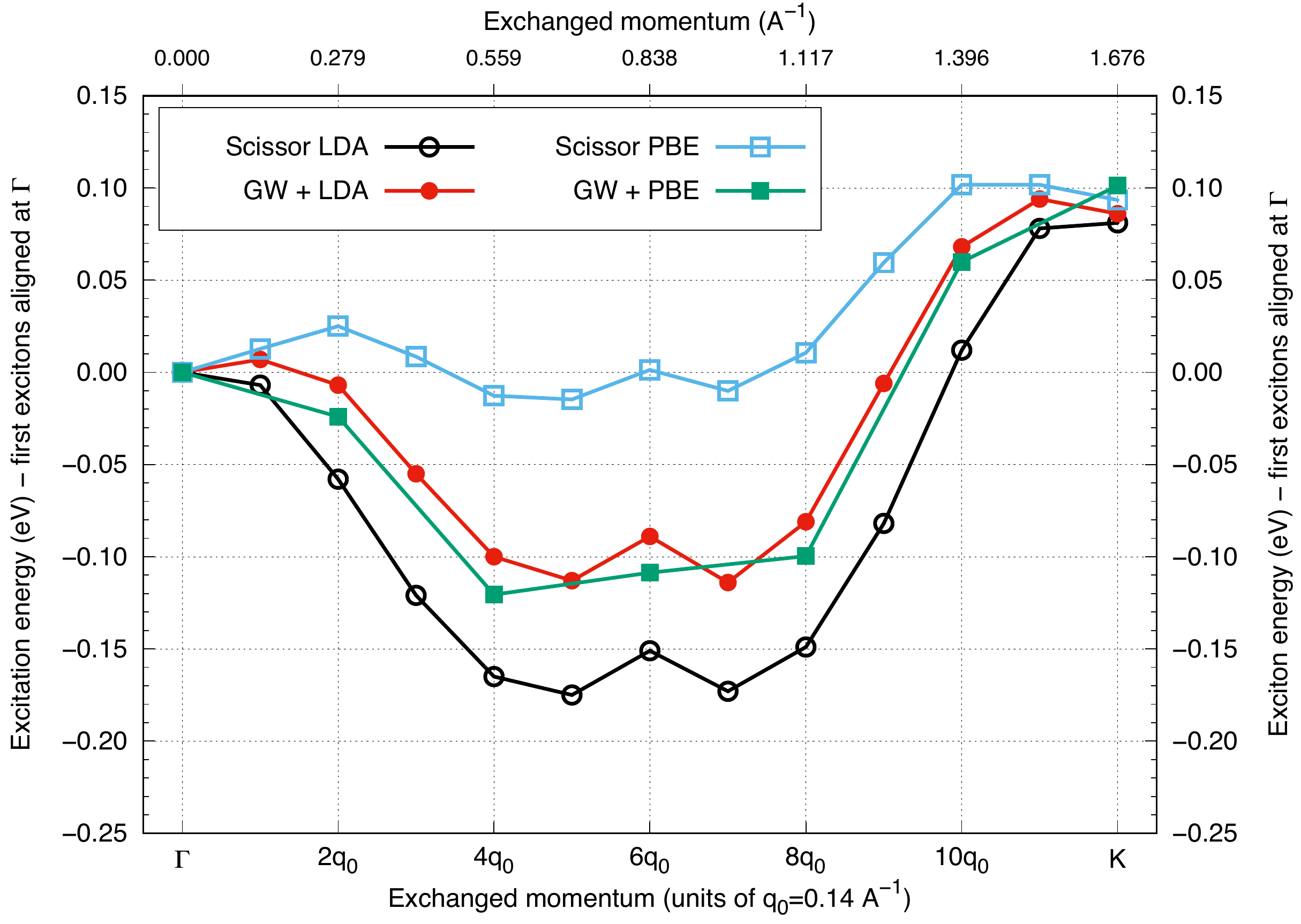}
\caption{Dispersion along $\Gamma K$ of the first exciton in AA' hBN computed with four different approximations. All curves have been aligned at $\Gamma$}
\label{fig:code_compar}
\end{figure}

Here we compare two calculations of the exciton dispersion in the AA' stacking done with different approximations for the quasiparticle correction and the DFT exchange-correlation potentials. The aim is clearly not to analyse the differences of the two approaches neither to make a thorough comparison of the two results. The objective is to assess the robustness of the results discussed in the main text.

In Figure~\ref{fig:code_compar} we report four dispersion curves of the first exciton in the AA' stacking for $\QQ \parallel \Gamma K$. All excitons have been aligned at $\Gamma$ to better visualise the variations on the energy dispersion.
Two calculations are based on LDA Kohn-Sham structure where the quasiparticle corrections have been approximated with a perturbative GW correction or a scissor operator. In the other two, the quasiparticle corrections have been computed within the same approximations but from PBE Kohn-Sham results.
The dispersion computed within the scissor operator are quite different, but still below 0.2~eV. However the two dispersions basically coincide when quasiparticle corrections are modelled within the GW approximation.

Furthermore it is remarkable that the two sets of calculations have been done with different codes. The LDA set of simulations have been obtained using the plane-wave codes EXC\cite{exc} and ABINIT,\cite{abinit} as described in~\ref{sec:methods:ab-initio:bulk}. The PBE set of simulations have been obtained using the GPAW\cite{Enkovaara2010} package. The very good agreement of the two GW results indicate that all parameters have been carefully converged in all calculations.

This comparison demonstrates the reliability and the robustness of our results, in particular regarding the claim of direct exciton formation in the AB bulk phase.

\bibliography{dispersion}

\end{document}